\documentclass[twoside,letterpaper]{emulateapj}
\pdfoutput=1
\usepackage{thumbpdf}
\usepackage{amsmath,amssymb}
\usepackage{graphicx,mathptm}
\usepackage{times}
\usepackage{natbib}
\usepackage[usenames]{color}
\usepackage{ifpdf}
\newcommand{\beq}{
\begin{equation}
}
\newcommand{\eeq}{
\end{equation}
}
\newcommand{\beqa}{
\begin{eqnarray}
}
\newcommand{\eeqa}{
\end{eqnarray}
}

\newcommand{\msun}     {\ensuremath{{\mathrm{M}}_{\scriptscriptstyle \odot}}}
\newcommand{\kms}      {\ensuremath{~\mathrm{km~s^{-1}}}}

\newcommand{\Mpc}      {\ensuremath{~\mathrm{Mpc}}}
\newcommand{\msigma}   {\ensuremath{M}{--}\ensuremath{\sigma}}
\newcommand{\ml}       {\ensuremath{M}{--}\ensuremath{L}}
\newcommand{\mbh}      {\ensuremath{M_{\mathrm{BH}}}}

\newcommand{\rinfres} {\ensuremath{R_{\mathrm{infl}} / r_{\mathrm{res}}}}
\newcommand{\figwidth}{0.475\textwidth}
\newcommand{\figwidthtwo}{0.35\textwidth}
\def\spose#1{\hbox to 0pt{#1\hss}}
\newcommand{\lta}{\mathrel{\spose{\lower 3pt\hbox{$\mathchar"218$}}
      \raise 2.0pt\hbox{$\mathchar"13C$}}}
\newcommand{\gta}{\mathrel{\spose{\lower 3pt\hbox{$\mathchar"218$}}
      \raise 2.0pt\hbox{$\mathchar"13E$}}}
\def\simlt{\mathrel{\rlap{\lower 3pt\hbox{$\sim$}}\raise 2.0pt\hbox{$<$}}}
\def\simgt{\mathrel{\rlap{\lower 3pt\hbox{$\sim$}} \raise 2.0pt\hbox{$>$}}}

\definecolor{KayhanCiteColor}{rgb}{0,0.08,0.35}
\definecolor{KayhanURLColor}{rgb}{0,0.08,0.35}
\definecolor{KayhanLinkColor}{rgb}{0,0.08,0.35}
\definecolor{KayhanPageColor}{rgb}{0,0.08,0.35}

\shorttitle{A Quintet of Black Hole Mass Determinations}
\shortauthors{G\"{u}ltekin et al.}

\makeatletter
\ifx\undefined\hyperref
\usepackage[pdftitle={A Quintet of Black Hole Mass Determinations},pdfauthor={Kayhan Gultekin},pdfsubject={Astrophysics},pdfkeywords={black hole physics --- galaxies: general --- galaxies:nuclei --- stellar dynamics},letterpaper,linktocpage,colorlinks=true,linkcolor={KayhanLinkColor},urlcolor={KayhanURLColor},citecolor={KayhanCiteColor}]{hyperref}
\else\relax\fi
\makeatother
\begin{document}

\label{firstpage}
 
\title{A Quintet of Black Hole Mass Determinations\footnotemark[1]}
\footnotetext[1]{Based on observations made with the \emph{Hubble
Space Telescope}, obtained at the Space Telescope Science Institute,
which is operated by the Association of Universities for Research in
Astronomy, Inc., under NASA contract NAS 5-26555.  These observations
are associated with GO proposals 5999, 6587, 6633, 7468, and 9107.}

\author{Kayhan
G\"{u}ltekin\altaffilmark{2}\altaffiltext{2}{Department of Astronomy,
University of Michigan, Ann Arbor, MI, 48109; Send correspondence
to \href{mailto:kayhan@umich.edu}{kayhan@umich.edu}.}, Douglas O. Richstone\altaffilmark{2}, Karl Gebhardt\altaffilmark{3}\altaffiltext{3}{Department of Astronomy, University of Texas, Austin,
TX, 78712.}, Tod R. Lauer\altaffilmark{4}\altaffiltext{4}{National Optical
Astronomy Observatory, Tucson, AZ 85726.}, Jason Pinkney\altaffilmark{5}\altaffiltext{5}{Department of
Physics and Astronomy, Ohio Northern University, Ada, OH 45810.}, 
%% \noaffiliation
%
M.~C. Aller\altaffilmark{6}\altaffiltext{6}{Department of
Physics, Institute of Astronomy, ETH Zurich, CH-8093 Zurich,
Switzerland.},
%% \noaffiliation
%
Ralf
Bender\altaffilmark{7}\altaffiltext{7}{Universitaets-Sternwarte der
Ludwig-Maximilians-Universit\"at, Scheinerstr. 1, D-81679 M\"unchen,
Germany.},
%% \noaffiliation
%
Alan Dressler\altaffilmark{8}\altaffiltext{8}{Observatories of
the Carnegie Institution of Washington, Pasadena, CA 91101.},
%% \noaffiliation
%
S.~M. Faber\altaffilmark{9}\altaffiltext{9}{University of
California Observatories/Lick Observatory, Board of Studies in
Astronomy and Astrophysics, University of California, Santa Cruz, CA
95064.},
%% \noaffiliation
%
Alexei
V. Filippenko\altaffilmark{10}\altaffiltext{10}{Department of
Astronomy, University of California, Berkeley, CA 94720-3411.},
%% \noaffiliation
%
Richard Green\altaffilmark{11}\altaffiltext{11}{LBT
Observatory, University of Arizona, Tucson, AZ 85721.},
%% \noaffiliation
%
Luis C. Ho\altaffilmark{8},
%% \noaffiliation
%
John Kormendy\altaffilmark{3},
%% \noaffiliation
%
Christos Siopis\altaffilmark{12}\altaffiltext{12}{Institut
d'Astronomie et d'Astrophysique, Universit\'{e} Libre de Bruxelles,
B-1050 Bruxelles, Belgium.}}
\affil{}

\begin{abstract}
%% %
We report five new measurements of central black hole masses based on
STIS and WFPC2 observations with the {\emph{Hubble Space Telescope}} and on 
axisymmetric, three-integral, Schwarzschild orbit-library kinematic models.  
We selected a sample of galaxies within a narrow range in velocity
dispersion that cover a range of galaxy parameters (including Hubble
type and core/power-law surface density profile) where we expected to
be able to resolve the galaxy's sphere of influence based on the
predicted value of the black hole mass from the \msigma\ relation.  We
find masses for the following galaxies:
NGC~3585,  $M_{\mathrm{BH}} = 3.4^{+1.5}_{-0.6}\times10^{8}~\msun$;
NGC~3607,  $M_{\mathrm{BH}} = 1.2^{+0.4}_{-0.4}\times10^{8}~\msun$;
NGC~4026,  $M_{\mathrm{BH}} = 2.1^{+0.7}_{-0.4}\times10^{8}~\msun$;
and NGC~5576, $M_{\mathrm{BH}} = 1.8^{+0.3}_{-0.4}\times10^{8}~\msun$,
all significantly excluding $\mbh = 0$.
For NGC~3945, $M_{\mathrm{BH}} = 9^{+17}_{-21}\times10^{6}~\msun$,
which is significantly below predictions from \msigma\ and \ml\
relations and consistent with $\mbh = 0$, though the presence of a
double bar in this galaxy may present problems for our axisymmetric
code.
\end{abstract}

\section{Introduction to Black Hole Mass Measurements}
\setcounter{footnote}{12}
\label{intro}

This paper is the latest in a campaign to model the central regions of
galaxies to determine masses of putative black holes 
from images and spectra taken
with the \emph{Hubble Space Telescope} (\emph{HST}).  The discovery of the
presence of a massive dark object (probably a black hole) in 
almost all galaxies having bulges, and
the scaling relations found versus host-galaxy properties, will stand as
one of the important legacies of this great observatory.

Masses were first determined from stellar velocity measurements made
with ground-based telescopes having the best possible spatial
resolution, together with isotropic kinematic models \citep{dr88}.
When combined with the spatial resolution of \emph{HST}, this method
has become the standard for black hole mass measurements
\citep[e.g.,][]{1998ApJ...493..613V,2000AJ....119.1157G}.  Black hole masses have also been derived
from stellar proper motions in our Galaxy
\citep{genzeletal00,2005ApJ...620..744G}, from megamaser measurements
of gas disks around central black holes
\citep[e.g.,][]{1995Natur.373..127M}, and from gas velocity
measurements \citep[e.g.,][]{barthetal01}.  Reverberation mapping has
also been used to find virial products in variable active galactic
nuclei \citep[AGNs; e.g.,][]{petersonetal04}.

Direct dynamical masses are the foundation for all scaling
relations used to infer black hole masses in active galaxies; all
measures of black hole mass are derived from the direct, dynamical
measurements.  Indirect mass indicators, such as AGN line widths, are
calibrated to reverberation mapping measurements \citep{bentzetal06},
which are themselves normalized against the direct dynamical
measurements \citep{onkenetal04}.

A central goal of a companion paper \citep{Gultekin_etal_2008b} is the
accurate measurement of the intrinsic or cosmic scatter in the
\msigma\ and $M$-$L$ relationships
\citep{magorrianetal98,gebhardtetal00}.  We have thus chosen to
augment the existing sample of \mbh\ measurements with new
determinations of $M_{\mathrm{BH}}$ for five galaxies selected to fall
within a narrow range in velocity dispersion.  Along with results from
the literature, this provides a number of galaxies in a narrow range
in velocity dispersion large enough that we may probe the intrinsic
scatter in the relation without biases incurred by, for example,
looking only at residuals to power-law fits.

We present observations of the centers of five early-type galaxies in
\S~\ref{obs}, including \emph{HST} observations in \S~\ref{stis},
ground-based imaging in \S~\ref{grimage}, and ground-based spectra in
\S~\ref{grspectra}.  We report results of dynamical models and black
hole masses in \S~\ref{model}, and summarize in \S~\ref{concl}.  In
the Appendix, we provide our data tables.

\section{Observations for New {$M_{\mathrm{BH}}$} Determinations}
\label{obs}
\subsection{Observational Sample}
\label{selection}

The five galaxies in this study were selected to come from a narrow
range in velocity dispersion ($180 < \sigma < 200\kms$) based on
HyperLEDA\footnote{Available at
\href{http://leda.univ-lyon1.fr/}{http://leda.univ-lyon1.fr/}.}
central velocity dispersion measures \citep{hyperleda}.  We chose this
range because (1) it includes galaxies with $M_B \approx -20$ mag
where both core and power-law surface-brightness profiles exist and
(2) it includes both late- and early-type galaxies.  We selected
galaxies with distances such that the predicted radius of influence
was larger than 0\farcs1.  The radius of influence is defined as
\beq
   R_{\mathrm{infl}} \equiv \frac{G \mbh}{\sigma^2\left(R_{\mathrm{infl}}\right)},
\label{e:rinfdef}
\eeq
where the velocity dispersion $\sigma$ is evaluated at the radius of
influence.  This obviously requires an iterative solution, but it
converges quickly.  The predicted mass comes from the central velocity
dispersion measurement and the \msigma\ fit of
\citet{tremaineetal02}.

The sample of new galaxies and their properties are presented in
Table~\ref{t:obssample}.  Distances are calculated assuming a Hubble
constant of $H_0 = 70~\kms~\mathrm{Mpc^{-1}}$.  We also provide
``Nuker Law'' surface-brightness profile parameters as a function of
radius given by
\beq I\left(r\right) = 2^{\left(\beta - \gamma\right)/\alpha} I_b 
   \left(\frac{r_b}{r}\right)^\gamma \left[1 + \left(\frac{r}{r_b}\right)^\alpha\right]^{\left(\gamma - \beta\right)/\alpha},
\label{e:nukerlaw}
\eeq
which is a broken power-law profile \citep{laueretal95}.  In addition
to the five galaxies whose black hole masses are reported in this
paper, we observed five others with \emph{HST} as part of the same observing
proposal.  Two of these (NGC~1374 and NGC~7213) had very low
signal-to-noise ratio (S/N).  The remaining three (NGC~2434, NGC~4382, and
NGC~7727) will be presented in a future paper.

\begin{deluxetable*}{llrrcrrrrr}
  \footnotesize
  \tablecaption{Observational Sample}
  \tablehead{
     \colhead{Galaxy} &
     \colhead{Type} &
     \colhead{Distance} &
     \colhead{$M_V$} &
     \colhead{Profile} &
     \colhead{$r_b$} &
     \colhead{$I_{b}$} &
     \colhead{$\alpha$} &
     \colhead{$\beta$} &
     \colhead{$\gamma$} 
  }
  \startdata
 N3585 &  S0 & 21.2 $\pm$ 1.8& $-22.01$ & $\wedge$   &  37.0 & 14.72 & 1.62 & 1.06 &  0.31\\
 N3607\tablenotemark{a} & E & 19.9 $\pm$ 1.6 & $-21.56$ & $\cap$ &  70.3 & 16.87 & 2.06 & 1.70 &  0.26\\
 N3945 & SB0 & 19.9 $\pm$ 3.0 & $-21.14$ & $\backslash$ &   3.9 & 18.62 & 0.30 & 2.56 & -0.06\\
 N4026 & S0 & 15.6 $\pm$ 2.0 & $-20.32$ & $\backslash$ &   3.0 & 15.23 & 0.39 & 1.78 &  0.15\\
%% N4382 & S0 & 17.9 & $\cap$     &  80.7 & 15.67 & 1.13 & 1.39 &  0.00\\
 N5576 &  E & 27.1 $\pm$ 1.7 & $-21.67$ & $\cap$     & 549.2 & 17.81 & 0.43 & 2.73 &  0.01
  \enddata
  \label{t:obssample}
  \tablenotetext{a}{NGC~3607 was listed incorrectly as being at a
  distance of $10.9~\mathrm{Mpc}$ by \protect{\citet{laueretal05}}.}
  \tablecomments{Distances are given in Mpc assuming a Hubble constant
  of $H_0 = 70~\kms~\mathrm{Mpc^{-1}}$.  All distances come from
  surface brightness fluctuations by \protect{\citet{tonryetal01}}
  except for NGC~3945, which comes from group distance by
  \protect{\citet{faberetal89}}.  The uncertainties to distances
  include random errors only.  The third column gives $V$-band
  absolute magnitudes taken from \protect{\citet{laueretal05}} and may
  be converted to $V$-band luminosities via $\log(L_V /
  {\mathrm{L}}_{\scriptscriptstyle \odot, V}) = 0.4 (4.83 - M_V)$
  \citep[see also][]{2008arXiv0807.1393V}.  The fourth column
  indicates surface-brightness profile type: power law ($\backslash$),
  core ($\cap$), or intermediate ($\wedge$) as determined by
  \protect{\citet{laueretal05}}.  ``Nuker-law'' surface-brightness
  profile parameters are given in columns 5 through 9 and correspond
  to Equation~\ref{e:nukerlaw}, where $r_b$ is the break radius in
  units of pc, $I_b$ is the surface brightness at the break radius in
  units of $V$ magnitudes per square arcsecond, $\alpha$ sets the
  sharpness of the profile break between the outer portion of the
  profile, which has power-law index of $\beta$, and the inner portion
  of the profile, which has power-law index of $\gamma$
  \protect{\citep{laueretal95,laueretal05}}.}
\end{deluxetable*}

\subsection{WFPC2 Imaging}
\label{spaceimage}

The high-resolution photometry of the central regions of the galaxies 
comes from Wide Field Planetary Camera 2 (WFPC2) observations on \emph{HST} 
using filters F555W ($V$) and F814W ($I$).  The observations, data
reduction, and surface-brightness profiles (including Nuker profile
fits) are detailed by \citet{laueretal05}. Surface-brightness profiles
are also available at the Nuker web page\footnote{See
\href{http://www.noao.edu/noao/staff/lauer/wfpc2\_profs/}{http://www.noao.edu/noao/staff/lauer/wfpc2\_profs/}.}.

\subsection{STIS Observations and Data Reduction}
\label{stis}
Our Space Telescope Imaging Spectrograph (STIS) observing strategy and 
data-reduction methods follow those of
\citet{pinkneyetal03}, which may be referred to for details.
Table~\ref{t:specs} gives the specifications for the STIS observations,
which used the G750M grating with either a $52\arcsec\times0\farcs1$ or
a $52\arcsec\times0\farcs2$ slit along the major axis of each galaxy and
the STIS 1024$\times$1024 pixel CCD with readout noise of $\sim 1
e^-$ at a gain of 1.0.  
For line-of-sight velocity distribution (LOSVD) fitting and for
measuring the STIS point-spread function (PSF), we used the previously
observed stellar spectral templates from \citet{pinkneyetal03} and
\citet{boweretal01}, which consist of a $V = 4.64~\mathrm{mag}$ G8~III
star (HR~6770), a $V = 5.03~\mathrm{mag}$ K3~III star (HR~7576), and a
$V = 3.909~\mathrm{mag}$ K0~III star (HR~7615).  The template stars
were scanned across the slit to mimic extended sources.

Most of the STIS setups used an unbinned CCD format with a read noise
of $\sim$ 1 $e^{-}$pixel$^{-1}$.  Wavelength range for all spectra was
8275-8847~\AA (Leitherer et al.\ 2001, pp. 231, 234). Reciprocal
dispersion measured using our own wavelength solutions was
0.554~\AA~pixel$^{-1}$.  The distribution of dispersion solutions for
a given data set had a $\sigma \approx 1.5\times10^{-4}$
\AA~pixel$^{-1}$.  The average dispersion given in the Handbook for
G750M is 0.56 \AA~pixel$^{-1}$.  We found a comparison line width of
$\sigma = 0.45~\AA = 17.5\kms$.  Instrumental line widths were
measured by fitting Gaussians to emission lines on comparison lamp
exposures.  This gives an estimate of the instrumental line width for
{\em extended} sources.  We use approximately five lines per exposure,
and at least five measurements per line.  While the comparison lamp
exposures were unbinned, the galaxy spectra were binned, which
increases the measured widths by $\sim$25\% for the 0\farcs1 slit and
by $\sim$ 3\% for the 0\farcs2 slit at 8561 \AA .  Leitherer et al.\
(2001, p. 300) give the following instrumental line widths for point
sources: $\sigma$=13.3, 15.0, 16.7\kms\ for the first three G750M
setups in Table~\ref{t:specs}.  The spatial scale
is
0\farcs05597 pixel$^{-1}$ for G750M at 8561 \AA.

\begin{deluxetable}{lllrccll}
\footnotesize
\tablecaption{STIS Long-Slit Spectrograph Configurations}
%% Turn on for emulateapj:
\tablewidth{0.4\textwidth}
\tablehead{
\colhead{}  &
\colhead{}   &
\colhead{Slit size}    &
\colhead{Exposure}    \\
\colhead{Name}          &
\colhead{Grating}    &
\colhead{$\arcsec \times \arcsec$ }    &
\colhead{s}    &
}
\startdata
NGC~3585 & G750M & 52$\times$0.1 & 12241 \\
NGC~3607 & G750M & 52$\times$0.2 & 26616 \\
NGC~3945 & G750M & 52$\times$0.2 & 22002 \\
NGC~4026 & G750M & 52$\times$0.1 &  9973 \\
NGC~5576 & G750M & 52$\times$0.1 &  7138
\enddata
\tablecomments{A summary of the main details of the STIS observational
set up.  Details can be found in the text.}

\label{t:specs}
\end{deluxetable}

The STIS data reduction was done with our own programs and FITSIO
subroutines \citep{pence98,pinkneyetal03}.  The raw spectra were
extracted from the multidimensional FITS file, and a constant fit was
subtracted from the overscan region to remove the bias level.  Because
the STIS CCD has warm and hot pixels that evolve on timescales of
$\la1$~d, the dark-current subtraction needs to be accurate.  We used
the iterative \emph{self-dark} method described by
\citet{pinkneyetal03}.  After flat fielding, we vertically shifted the
spectra to a common dither, combined them, and then rotated.
One-dimensional spectra were extracted from the final spectrogram
using a biweight combination of rows.  A 1-pixel wide binning scheme
was used near the galaxy center to optimize spatial resolution.  We
present Gauss--Hermite moments of the velocity profiles in the
Appendix.

\subsection{Ground-Based Imaging}
\label{grimage}

CCD images of one of our galaxies, NGC~3945, were obtained from the
MDM 1.3-m McGraw-Hill Telescope.
The $2048 \times 2048$ pixel CCD named {\it Echelle} was used.  This chip has
0.508$''$ pixel$^{-1}$ and a readout noise of 2.7 e$^{-}$ pixel$^{-1}$.
The conditions were clear but not reliably photometric at all times,
and the seeing varied between 1\farcs6 and 2\farcs4.  The combined
images all had a full width at half-maximum intensity (FWHM) 
of 2\farcs0$\pm 0\farcs1$.  (The surface-brightness 
profiles at small radii are taken from \emph{HST} data, so the
relatively large FWHM is not a serious problem.)  NGC~3945 was
observed in $I$, $V$, and $R$ filters.  Standard CCD reduction tasks
within IRAF were used to subtract overscan, trim overscan, and divide
by flat-field frames.  The task {\tt cosmicrays} was used to remove
cosmic rays because the number of exposures was too small for median
filtering to work with the $V$ and $R$ bands.  The $I$ band suffers
from interference fringes which did not flatten out.  The
flat fielding was good to 1\% of the sky background in $R$, 2\% in
$I$, and 0.6\% in $V$.  We present the $V$-band surface-brightness
profile for NGC~3945 profile in the Appendix.

We also obtained wide-field $V$-band surface-brightness profiles from the
following sources in the literature: NGC~3585 is from \citet{bsg94}; 
while NGC~3607,
NGC~4026,
 and
NGC~5576 are from \citet{mm93}.

\subsection{Ground-Based Spectra}
\label{grspectra}

Absorption-line spectra were obtained from three ground-based
telescopes in order to derive stellar kinematics out to large
radii.  Table~\ref{t:gndspecs} describes the spectrograph setups.
The instrumental resolution, as estimated from widths of comparison 
lines, was below 60 \kms\ in all cases. This allows the galaxy
line widths ($\sigma \approx 200$ \kms ) to be easily resolved. 
Using the Modular Spectrograph (ModSpec) at MDM Observatory
with the Wilbur CCD we are able to observe the near-infrared
Ca II triplet without fringing.  At Magellan, however, fringing
was a problem and so the Mg b spectral range was used.
Our seeing estimates came from consecutive star observations using 
the same setup. These were confirmed by a seeing monitor
in the case of Magellan.

\begin{deluxetable}{llll}
\tablecaption{\sc Ground-based Spectrographs} 
\tablehead{
\colhead{}   & 
\colhead{MDM} & 
\colhead{Magellan I} & 
\colhead{Magellan II} \\ 
\colhead{}   &
\colhead{2.4-m}      &
\colhead{6.5-m}      &
\colhead{6.5-m}   
}
\startdata
Spectrograph  & ModSpec       & B\&C           & B\&C  \nl
CCD           & Wilbur        & Tek 1         & Marconi I  \nl
Central $\lambda$ & $\sim $ 8500 \AA & $\sim $ 5175 \AA & $\sim 5175 $ \AA \nl
                  & Ca II            & Mg b             & Mg b  \nl
Line widths ($\sigma$)\tablenotemark{a} & 1.1 \AA & 0.97\AA  & 0.95 \AA \nl
                                        & 39~\kms & 56~\kms  & 55~\kms \nl
Dispersion (\AA\ pixel$^{-1}$) &  0.9   &  1.4    &  0.78   \nl
Slit length\tablenotemark{b} &  500\arcsec & 70\arcsec & 60\arcsec    \nl
Seeing (FWHM) & 1.3\arcsec   & 0.65\arcsec   & 0.6\arcsec  \nl
Slit Width  & 0.8\arcsec    & 0.71\arcsec   & 0.71\arcsec  \nl
Spatial Scale & 0.371\arcsec pixel$^{-1}$ & 0.44\arcsec pixel$^{-1}$   &  0.25\arcsec pixel$^{-1}$  %\nl \hline
\enddata
\tablenotetext{a}{Measured from comparison lamp emission lines.}
\tablenotetext{b}{As limited by CCD format.}
\label{t:gndspecs}
\end{deluxetable}

Table~\ref{t:gndobs} summarizes the observations of NGC~3945 and
NGC~5576.  The same, basic observing procedure was used at Magellan
and MDM.  At the beginning of the run, the slit width was set to
values typical of the seeing, $\sim$ 0\farcs7--1\farcs0.  Bias frames,
continuum lamp flats, twilight sky flats, and comparison lamp spectra
were taken before and/or after the night.  Calibration frames were also
taken consecutively with the galaxies.  These included comparison
lamps, template stars, and focus stars.  At MDM, the focus frames were
created by moving the star to new positions along the slit during the
pauses between 5 and 7 subexposures.  These frames allowed us to model
the spatial distortions in the galaxy exposures more precisely.  At
Magellan, we only had the galaxy peaks and single-star exposures to
define the ``S-distortion."  During subsequent runs, however, we used
a flat with a multislit decker to see that the S-distortion of a star
near the edge of the slit would be the same as the S-distortion at the
center within one-pixel width.  We estimate the spatial distortions
should be less than $\sim 0\farcs5$ from center to edge.

For each galaxy, we obtained at least two slit position angles (P.A.s)
to improve spatial coverage.  The targets were observed within
$|\mathrm{HA}| < 2~\mathrm{hr}$ to minimize atmospheric refraction and
extinction.  Multiple galaxy exposures were made at each slit position
to improve the S/N and to median-filter cosmic rays.  Some dithering
was employed to lessen the impact of chip defects.  We first obtained
exposures at the major axis slit position.  After two to five
exposures, we started the rotated exposures.  For Magellan~I with
Tek1, a 600~s exposure gave us S/N = 50 per \AA\ per 1\arcsec\ wide
bin near the Mg~b line at the center of the galaxy. Only two exposures
were required for good S/N even at larger radii (where many rows were
binned).  For MDM, a similar (600 s) exposure gave only S/N $\approx$
18.4 per 1\AA\ per 1\arcsec\ bin.  We therefore used 1200~s exposures
(S/N $\approx$ 25.9) at MDM and aimed for more exposures.

\begin{deluxetable}{lllrrl}
\tablecaption{\sc Ground-based Observations} 
\tablehead{
\colhead{Name} & 
\colhead{} & 
\colhead{} & 
\colhead{P.A. \tablenotemark{c} } & 
\colhead{Exposure \tablenotemark{d}}  &
\colhead{Type of \tablenotemark{e}}  \\
\colhead{NGC}     & 
\colhead{Date \tablenotemark{a}} & 
\colhead{Telescope \tablenotemark{b}} & 
\colhead{ \arcdeg } & 
\colhead{($\mathrm{s}$) } & 
\colhead{Observation}   
}
\startdata
3945 & 3/16/01 & MDM 2.4 m & 0 & 3$\times$1200 & CaT  \nl  % R2N5
3945  & 3/16/01 & MDM 2.4 m & 90 & 3$\times$1200 & CaT  \nl  % R2N5
3945  & 3/17/01 & MDM 2.4 m & 90 & 2$\times$1200 & CaT  \nl  % R2N6
3945  & 5/10/03 & MDM 2.4 m & 0 & 3$\times$1200 & CaT  \nl  % A3N6
3945  & 3/15/01 &MDM 1.3 m & --- & 3$\times$300 & $I$ image  \nl % R3N2
3945  & 3/15/01 &MDM 1.3 m & --- & 2$\times$300 & $V$ image  \nl % R3N2
3945  & 3/15/01 &MDM 1.3 m & --- & 1$\times$300 & $R$ image  \nl %%\hline % R3N2
\\
5576  & 6/22/01 & Magel1 6.5 m & 0 & 1$\times$600 & Mg~b \nl  % D4N4
5576  & 6/22/01 & Magel1 6.5 m & 60 & 2$\times$600 & Mg~b \nl  % D4N4
5576  & 4/07/03 & Magel2 6.5 m & 0 & 2$\times$1200 & Mg~b  %\nl \hline % D7N3
\enddata
\tablenotetext{a}{The date of observation given as MM/DD/YY. }
\tablenotetext{b}{The observatory and telescope. }
\tablenotetext{c}{The position angle of the slit relative to
the major axis of the galaxy. }
\tablenotetext{d}{Number of exposures $\times$ exposure length (s).}
\tablenotetext{e}{Type of observation.  CaT = includes Ca II triplet near 8500 \AA ,
Mg~b = includes Mg~b feature near 5175 \AA .}
\label{t:gndobs}
\end{deluxetable}

The two-dimensional (2D) spectra were reduced using the IRAF tasks
found primarily in the {\tt ccdred}, and {\tt twodspec} packages.
Bias subtraction was generally not important, but we
overscan-corrected and trimmed the CCD frames.  Flat-fielding was
performed using frames constructed out of three flats. First, a
twilight flat was used to define the illumination pattern of sky light
along the slit. Second, the small-scale structure flat was created by
dividing the continuum lamp by a smooth fit.  Third, the large-scale
structure along the dispersion axis was defined in a flat created by
fitting 1D polynomials to each row of the continuum lamp flat.  These
three normalized flats were multiplied to obtain the final flat.
Since the continuum lamp does not have a perfectly flat spectrum,
flat-fielding does not produce an accurate galaxy continuum.  However,
the galaxy spectra are normalized before LOSVDs are drawn, so this
does not significantly interfere with our kinematics.

After flat fielding, the spectra were wavelength-calibrated and
corrected for spatial distortions.

This requires first finding wavelength calibrations as a function of
position along the slit.  We used Ar, Ne, and He comparison lamp
exposures to define wavelength as a function of position.  The fit
along dispersion axis was typically a fourth-order Legendre polynomial.
This provided a root mean square (rms) residual of $\sim$0.15~\AA .
As discussed above, the spatial axis was rectified using the peaks of
stars and galaxies.

The only remaining steps in reduction were sky subtraction and
combining exposures.  The sky subtraction was performed by subtracting
a fit to the counts along each cross-dispersion row with a low-order
polynomial (usually a constant or a line) but excluding the central
pixels containing significant galaxy light.  This method worked well
for the MDM data with its longer slit.  Fortunately, for the Magellan
I NGC~5576 observation, the Moon was down and the sky does not
seriously affect our line strengths even if it is not subtracted.  For the
Magellan II observations of NGC~5576, other program objects with a
smaller spatial extent were used to produce a sky spectrum.  Finally,
we averaged exposures using a cosmic-ray rejection option.  We only
combined exposures if they were taken on the same night and with the
same slit position.  Any dithering between exposures was removed by
shifting all galaxy peaks to a common row or column.  We present
Gauss--Hermite moments of the velocity profiles in the Appendix.

\subsection{LOSVDs}
\label{losvds}
\begin{figure}
%% \pdfbookmark[1]{Fig. 1: NGC 3585 velocity profile}{pbf1}
\centering
\includegraphics[width=\figwidth]{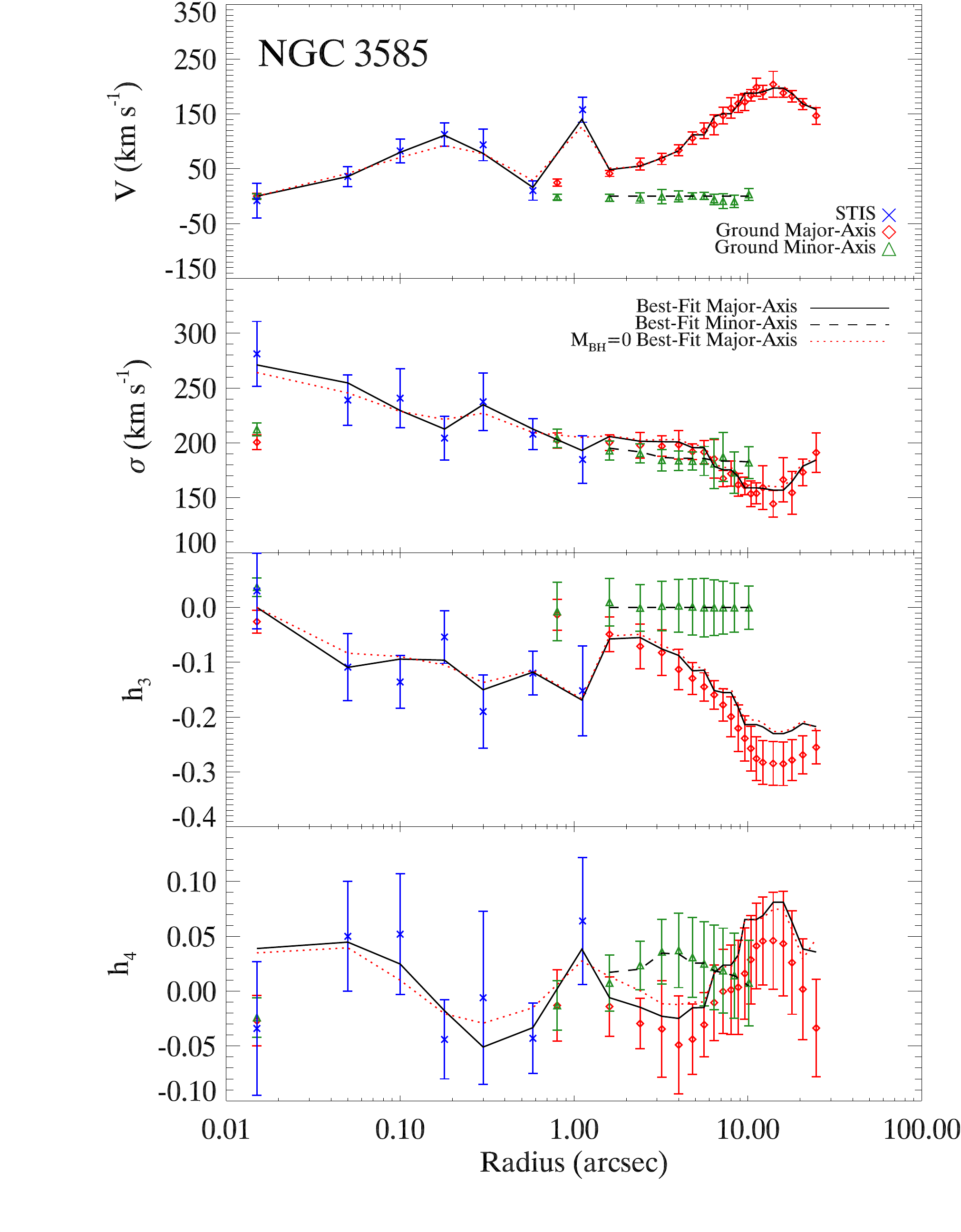}
\caption{Gauss--Hermite moments of LOSVDs for NGC~3585. Blue crosses
are Gauss--Hermite moments of LOSVDs from \emph{HST} STIS data.  Also
plotted are ground-based Gauss--Hermite moments of LOSVDs along the
major axis (red diamonds) and minor axis (green triangles) from
\protect{\citet{fisher97}}.  \protect{\citet{fisher97}} binned $v$ and
$\sigma$ differently from the third and fourth moments.  We
interpolated and rebinned them consistently.  Because the ground $h_3$
and $h_4$ moments have been interpolated, their error bars are larger
than their scatter.  Though Gauss--Hermite moments are not fitted directly
in the modeling, the jagged black lines are the resulting
Gauss--Hermite fit to the best-fit model's LOSVDs from
\S~\protect{\ref{model}} for the major axis (solid black) and minor
axis (dashed black).  The best-fit model has $M_{\mathrm{BH}} =
3.4\times10^8~\msun$ and $\Upsilon = 3.5$.  The STIS spectra show a
rise in velocity dispersion toward the center and require a black hole
to match the increased mass-to-light ratio.  The red dotted line shows the
best-fit model for which $M_{\mathrm{BH}} = 0$, which has $\Upsilon =
4.1$.}
\label{f:n3585spec}
\end{figure}

\begin{figure}
\centering
%% \pdfbookmark[3]{Fig. 2: NGC 3607 velocity profile}{pbf2}
\includegraphics[width=\figwidth]{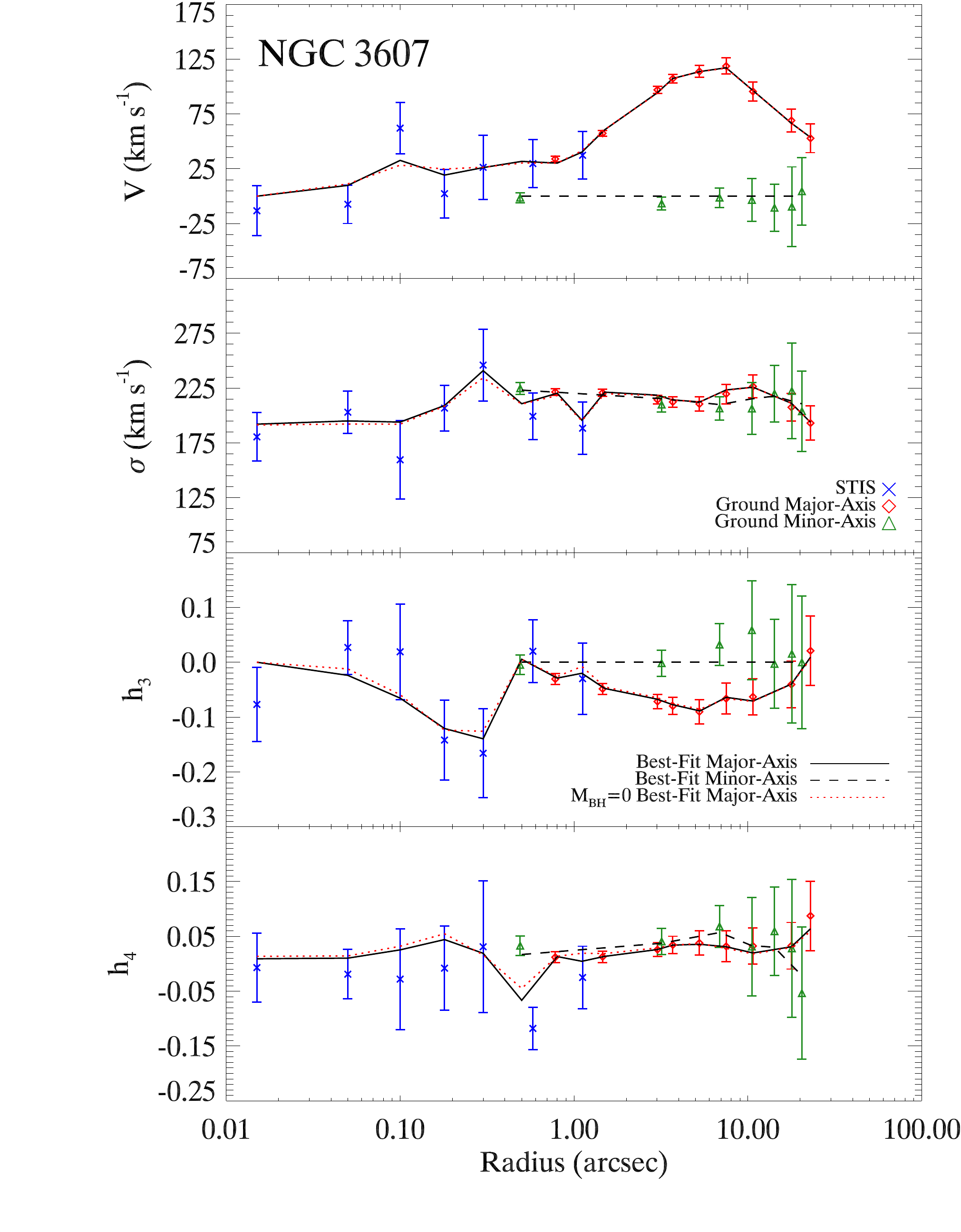}
\caption{Gauss--Hermite moments of LOSVDs for NGC~3607.  Symbols are as
in Figure~\protect{\ref{f:n3585spec}}.  Ground-based data are from
\protect{\citet{bsg94}}.  The best-fit model (black lines) has
$M_{\mathrm{BH}} = 1.25\times10^8~\msun$ and $\Upsilon = 7.3$.  The
best-fit model without a black hole (red, dotted line) has $\Upsilon =
7.5$.}
\label{f:n3607spec}
\end{figure}

\begin{figure}
\centering
%% \pdfbookmark[3]{Fig. 3: NGC 3945 velocity profile}{pbf3}
\includegraphics[width=\figwidth]{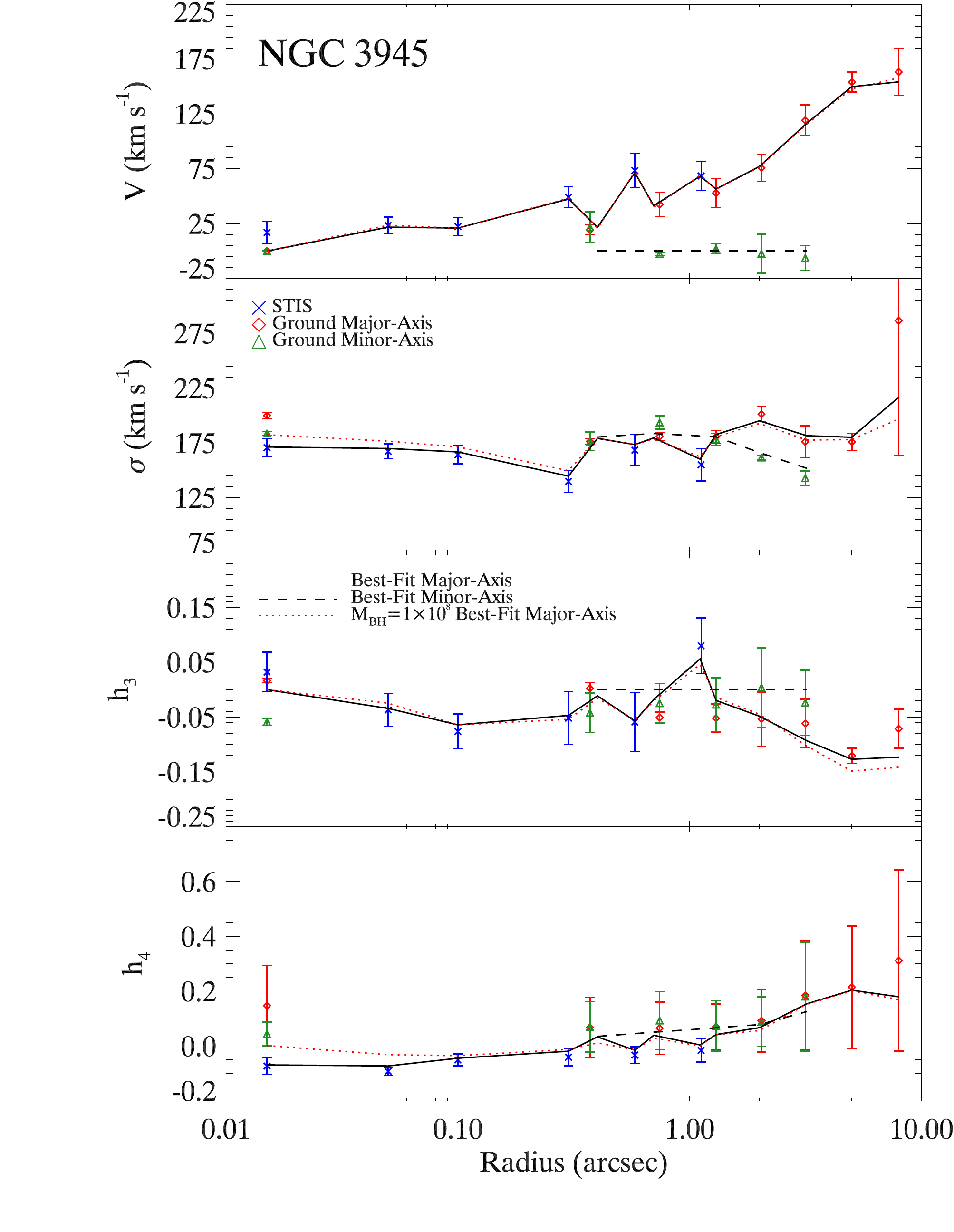}
\caption{Gauss--Hermite moments of LOSVDs for NGC~3945.  Symbols are as
in Figure~\protect{\ref{f:n3585spec}}.  Ground data are from our MDM
observations.  Error bars for ground-based data are from variations
from one side of the galaxy to the other, which dominate the total
uncertainty for this galaxy.  The black jagged lines are from the
best-fit model, which has $M_{\mathrm{BH}} = 0$ and $\Upsilon = 7.2$.}
\label{f:n3945spec}
\end{figure}

\begin{figure}
\centering
%% \pdfbookmark[3]{Fig. 4: NGC 4026 velocity profile}{pbf4}
\includegraphics[width=\figwidth]{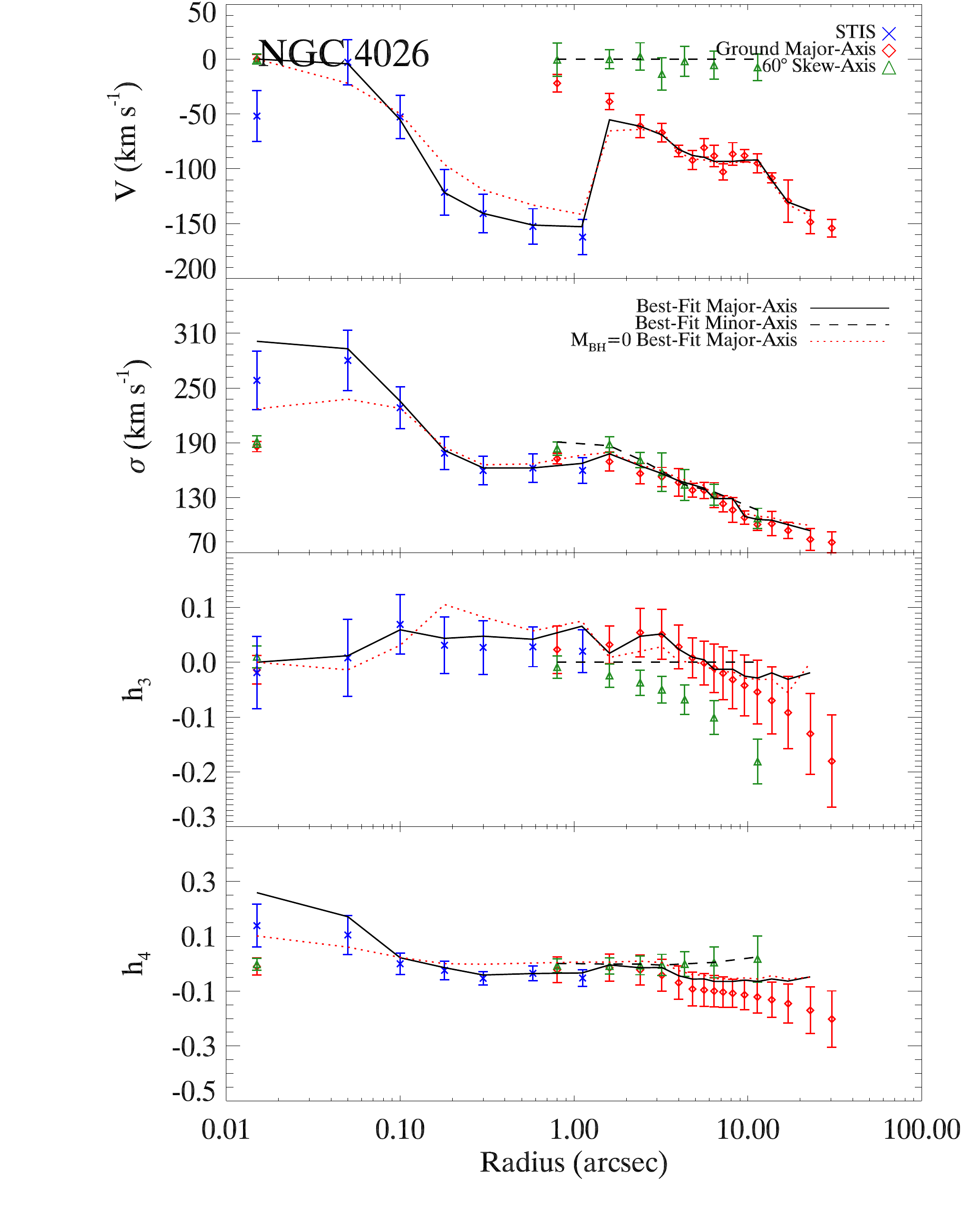}
\caption{Gauss--Hermite moments of LOSVDs for NGC~4026.  Symbols are as
in Figure~\protect{\ref{f:n3585spec}}.  Ground-based data are from
\protect{\citet{fisher97}}, for which the $h_3$ and $h_4$ moments have been
interpolated.  Because of the interpolation, the scatter in the data
is less than the error bars.  The LOSVDs show a sharp increase in
velocity dispersion toward the center.  The jagged lines are from the
best-fit model, which has $M_{\mathrm{BH}} = 2.2\times10^8~\msun$ and
$\Upsilon = 4.6$.  The best-fit model without a black hole (red dotted
line) has $\Upsilon = 5.6$.}
\label{f:n4026spec}
\end{figure}

\begin{figure}
\centering
%% \pdfbookmark[3]{Fig. 5: NGC 5576 velocity profile}{pbf5}
\includegraphics[width=\figwidth]{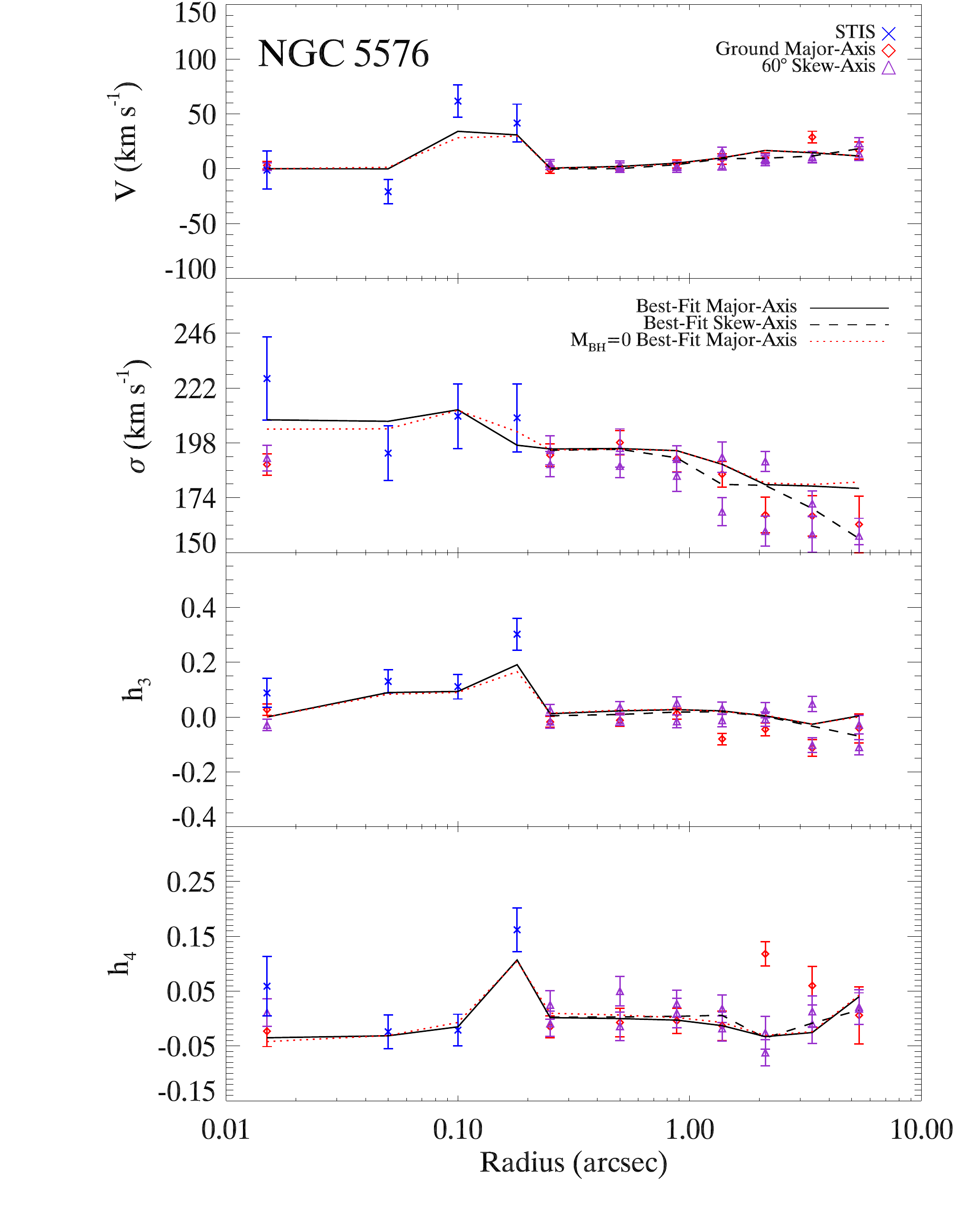}
\caption{Gauss--Hermite moments of LOSVDs for NGC~5576.  Symbols are as
in Figure~\protect{\ref{f:n3585spec}}.  The ground-based data are from
our Magellan observations.  The jagged lines are from the best-fit
model, which has $M_{\mathrm{BH}} = 1.6\times10^8~\msun$ and $\Upsilon
= 3.6$.  The best-fit model without a black hole (red dotted line) has
$\Upsilon = 4.0$.}
\label{f:n5576spec}
\end{figure}

Our modeling of stellar kinematics is done by comparing binned
LOSVDs of our models to those
derived from the galaxy spectra.  LOSVDs are calculated at the
positions indicated in Figures~\ref{f:n3585spec}--\ref{f:n5576spec}.
The STIS spectra probe the inner 1\farcs1, and the ground-based
spectra probe the outer regions.  We combine the LOSVDs extracted from
both sets of data to give us kinematic descriptions of the galaxies
from both inner and wide-field regions.  We deconvolve the
observed galaxy spectrum using the template spectrum composed from the
standard stellar spectra.  The deconvolution is done with the maximum
penalized-likelihood method described by \citet{gebhardtetal03} and
\citet{pinkneyetal03}.

Ground-based data obtained from the literature in the form of
Gauss--Hermite moments with associated errors were converted to LOSVDs
using Monte Carlo simulations to estimate the uncertainties in the
velocity profile bins.  The Monte Carlo simulations used $10^4$
realizations for each LOSVD to sample the uncertainties.  If the
Gauss--Hermite moments corresponded to an unphysical negative value for
the LOSVD, we assigned it a value of zero with a conservative
uncertainty.  Based on previous experience, we bin the velocity
profile into 13 equal bins that cover the range of velocities seen in
the given galaxy.  Gauss--Hermite data from opposite sides of the
galaxy were typically averaged (changing the sign of odd moments)
because our models are axisymmetric.  For NGC~5576, however, we used
the LOSVDs from both sides of the galaxy independently, changing the
sign of the velocity so as to be used with our axisymmetric model.
The kinematic data from \citet{fisher97} binned the higher-order
moments ($h_3$ and $h_4$) differently from the lower-order moments ($V$
and $\sigma$).  We interpolated and rebinned them consistently.  The
smoothness of the data suggests that this does not introduce a large
systematic error.

%%%%%%%%%%%%%%%%%%%%%%%%%%%%%%%

\newpage
\section{Dynamical Models to Estimate {$M_{\mathrm{BH}}$}}
\label{model}
In this section we present the results of models to test for the
presence of a central black hole.  We use the three-integral,
axisymmetric Schwarzschild method to make dynamical models of the
galaxies.  The method is explained by \citet{gebhardtetal03} and in
more detail by \citet{siopisetal08}, but we very briefly outline it
here. 

First, we use the photometric data with the assumption of
axisymmetry and a given inclination to find the luminosity density of
the galaxy.  We use the luminosity density with a given mass-to-light
ratio ($\Upsilon$) and a given black hole mass to calculate the potential.  We then
calculate the orbits of representative stars in this potential.  From
this, we determine the weights for the set of orbits that best
reproduces the surface-brightness profile.  With the weighted orbit
library, we find the LOSVD for a
given inclination for comparison with the observed spectra.  

The results are summarized in Table~\ref{t:masses}.  For all galaxies,
we modeled several inclination angles, but in all cases none was
clearly preferred by the models alone, and all found the same black
hole mass within the stated uncertainties.  We present the values from
our edge-on models below except for NGC~3607, whose image indicates that
it is face-on.  Measurement errors are 1$\sigma$ uncertainties.  We
generally report two values for black hole mass and mass-to-light
ratio in the text: (1) the best-fit-model value which comes from the
single model with the smallest $\chi^2$ and (2) the value obtained
from marginalizing over the other parameter.  These two values are not
always exactly equal, but the best-fit model is always within
1$\sigma$ of the marginalized value.  We put the marginalized values,
which incorporate our uncertainty in the other parameter, in
Table~\ref{t:masses} and use those values for all subsequent
calculations.

We have investigated the reliability of this orbit-superposition
program in a number of ways.  The results from the program have been
compared to those from the Leiden program developed by McDermid et
al. (in preparation), giving consistent results for NGC~0821 using the
different codes on the same data and on different data of the same
galaxy.  \citet{siopisetal08} also tested
our method by synthesizing a distribution function for a galaxy and
running them through the modeling program.  The models were able to
recover the black hole mass as well as the orbital configuration.

\citet{gebhardt04} 
demonstrated that a sufficiently large orbit library produces
consistent results with other orbit libraries of similar or larger
sizes.  He showed that the black hole mass is not influenced by the
choice of the weight on entropy in the solution process, provided it
is sufficiently small.  \citet{gebhardtetal03} also showed that a set
of objects observed at \emph{HST} resolution and ground-based
resolution gives consistent (but with different precisions) black hole
masses when only the ground-based data are used to construct models.
\citet{kormendy04}
showed that estimates of the mass of the black hole in M32 using
techniques similar to these but different in detail give results
consistent with the current value (and their own error bars) over a
10-fold improvement in spatial resolution.  Hence, in the context of
these models we believe the program returns correct estimates of black
hole mass and mass-to-light ratio, and we adopt these estimates below,
even when the radius of influence of the resulting black hole,
$R_{\mathrm{infl}} = G \mbh / \sigma^2(R_\mathrm{infl})$, is less than
the resolution of the observation.

There are at least three issues that might lead us to report black
hole masses that are \emph{significantly} wrong (i.e., outside our
error bars).  First, the models are axisymmetric by construction,
hence significant triaxiality could lead to an error.  Triaxial
models, of course, can reconstruct the correct structure
\citep[e.g.,][]{vandenboschetal08}.  Second, the models are assumed to
have constant stellar mass-to-light ratios except for the central
black hole.  An admixture of dark matter with a spatial distribution
different from that of the luminous matter could lead us to determine
an incorrect black hole mass thus requiring dark matter in the model
\citep[e.g.,][]{thomasetal07}.  Finally, \citet{houghtonetal06} have
argued for methods of determining the LOSVD that address the
limitation of our method (maximum penalized likelihood)---that it
produces LOSVDs with correlated errors.  \citet{gebhardtetal03} did
Monte Carlo simulations that give us confidence in our ability to
estimate the number of independent data in our LOSVDs.  Nonetheless,
our method could be improved.

\begin{deluxetable*}{lrrrrlrr}
  \footnotesize
  \tablecaption{Mass Measurements}
  \tablehead{
     \colhead{Galaxy} &
     \colhead{$\sigma_e$} &
     \colhead{$M_{\mathrm{BH}}$} &
     \colhead{$M_{\mathrm{BH,low}}$} &
     \colhead{$M_{\mathrm{BH,high}}$} &
     \colhead{$\Upsilon_V$} &
     \colhead{$\chi^2$} &
     \colhead{$\Delta\chi^2$} 
  }
  \startdata
  NGC~3585 & 213 &  $3.4\times10^{8}$ & $2.8 \times 10^8$ & $4.9 \times 10^8$ & 3.4  $\pm$ 0.2  & 55.8 & 28.7\\
  NGC~3607 & 229 &  $1.2\times10^{8}$ & $7.9 \times 10^7$ & $1.6 \times 10^8$ & 7.5  $\pm$ 0.3  & 69.1 & 10.6\\
  NGC~3945 & 192 &  $9\times10^{6}$  & $-1.2 \times 10^7$ & $2.6 \times 10^7$ & 6.6  $\pm$ 0.8  & 46.2 & . . . .\\
  NGC~4026 & 180 &  $2.1\times10^{8}$ & $1.7 \times 10^8$ & $2.8 \times 10^8$ & 4.5  $\pm$ 0.3  & 78.7 & 26.2\\
  NGC~5576 & 183 &  $1.8\times10^{8}$ & $1.4 \times 10^8$ & $2.1 \times 10^8$ & 3.7  $\pm$ 0.3  & 319.1 & 15.5\\
  \enddata
  \label{t:masses}
  \tablecomments{Results from mass modeling.  Effective stellar
  velocity dispersions are given in units of $\kms$, masses are in
  units of $\msun$, and $\Upsilon_V$ is units of
  $\msun~{\mathrm{L}}^{-1}_{\scriptscriptstyle \odot,V}$.  The black hole masses and mass-to-light
  ratios are the result of marginalizing over the other parameter.
  $M_{\mathrm{BH,low}}$ and $M_{\mathrm{BH,high}}$ are the 1$\sigma$
  confidence limits on the detected black hole mass.  The final two
  columns list $\chi^2$ of the best-fit model, and the difference
  between the minimum in the marginalized $\chi^2$ and at 
  $M_{\mathrm{BH}} = 0$.}
\end{deluxetable*}

\subsection{NGC~3585}
\label{n3585}
\object[NGC 3585]{NGC~3585} is an edge-on S0 galaxy at a distance of $21.2 \Mpc$
\citep{rc3,tonryetal01}.  Images of the galaxy show that it is
flattened with a nearly edge-on dust ring at its center.  The LOSVD
profile shows a roughly constant velocity dispersion from about
$1\arcsec$ to $\sim6\arcsec$ of $\sigma \approx 200\kms$.  Inside
$\sim0\farcs1$ the dispersion rises to $\sim280\kms$, indicating a
likely increase of mass-to-light ratio toward the center.  For use in
analysis of the \msigma\ relation, we compute an effective stellar
velocity dispersion,
\beq
\sigma^2_e \equiv \frac{\int_{0}^{R_e} \left({\sigma^2 + V^2}\right) I\left(r\right) dr}{{\int_{0}^{R_e} I\left(r\right) dr}},
\label{e:sigmae}
\eeq
where $R_e$ is the effective radius, $I(r)$ is the surface-brightness
profile, and $V$ and $\sigma$ are the first and second Gauss--Hermite
moments of the LOSVD from a slit of width 1\arcsec.  From the ground-based
velocity profile, we find an effective stellar velocity dispersion of
$\sigma_e = 213\kms$.

\begin{figure}
\centering
%% \pdfbookmark[3]{Fig. 6: NGC 3585 chi-square contours}{pbf6}
\includegraphics[width=\figwidthtwo,angle=90]{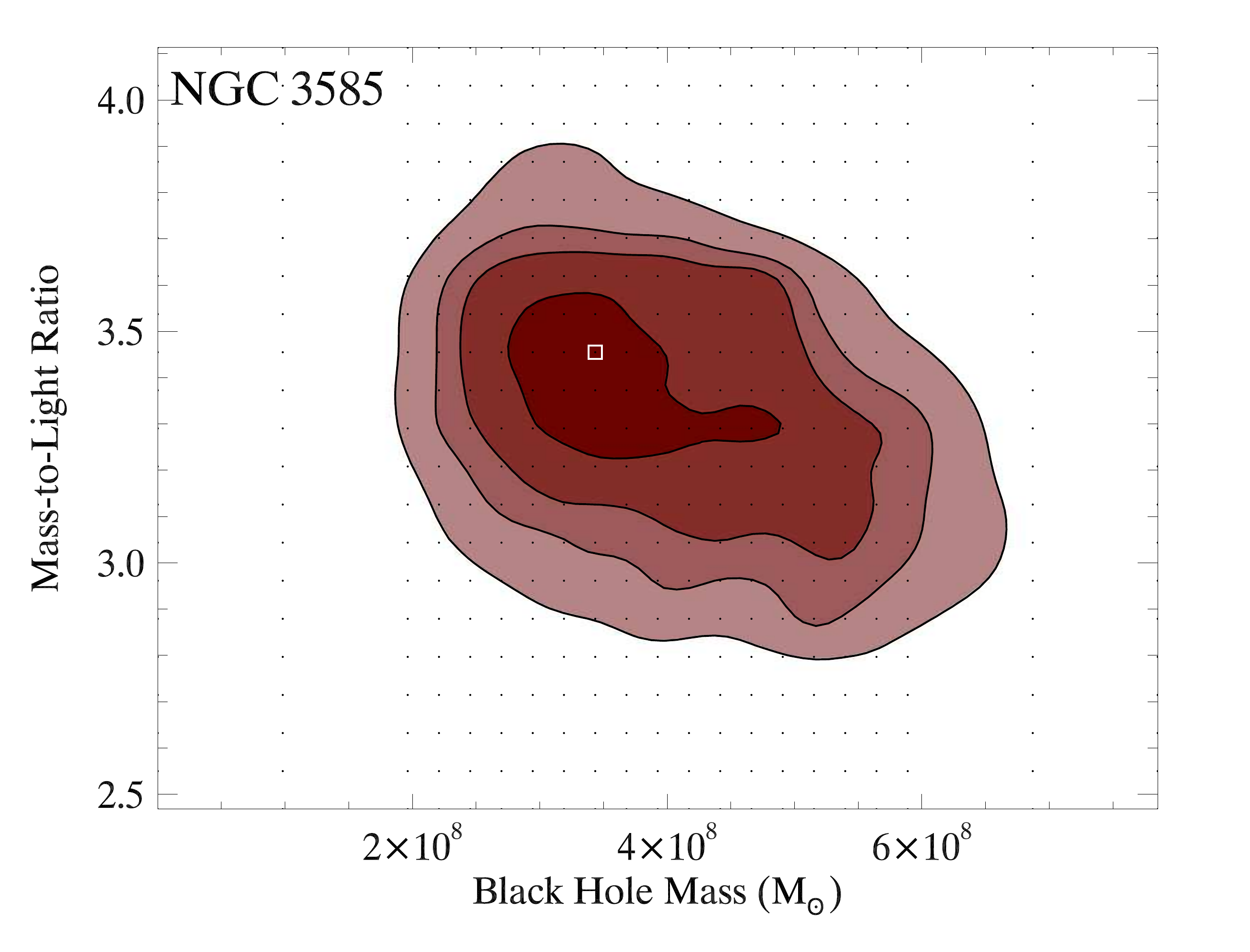}
\caption{Mass modeling $\chi^2$ contours for NGC~3585, assuming
edge-on inclination.  Contours are for $\Delta\chi^2 =$ 1.00, 2.71,
4.00, and 6.63, which bracket individual parameter confidence levels
of 68.3, 90.0, 95.4, and 99.0\%, respectively.  Contours have been
smoothed for plotting.  The square shows the best-fit model.  Dots
indicate parameters modeled.  The contours were smoothed for plotting.
The best-fit model has $M_{\mathrm{BH}} = 3.4\times10^8~\msun$ and
$\Upsilon = 3.5$.  Marginalizing over the other parameter we find
$M_{\mathrm{BH}} = 3.4^{+2.5}_{-0.6}\times10^8~\msun$ and $\Upsilon =
3.4\pm 0.2$.}
\label{f:n3585chi}
\end{figure}

\begin{figure}
\centering
%% \pdfbookmark[3]{Fig. 6: NGC 3585 cumulative chi-square difference}{pbf7}
\includegraphics[width=\figwidth]{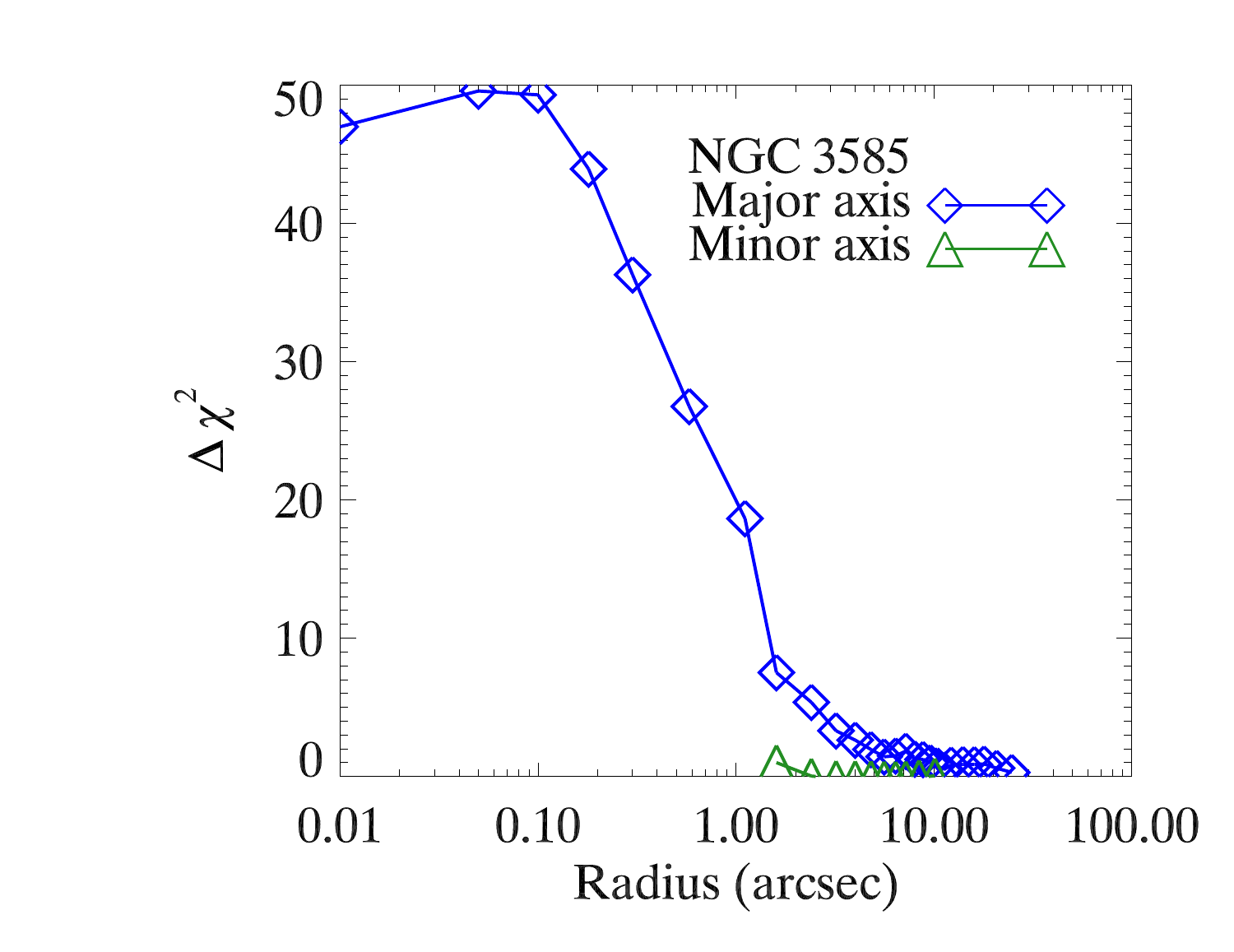}
\caption{Cumulative $\chi^2$ difference between the best-fit model
  ($M_{\mathrm{BH}} = 3.4\times10^8~\msun$ and $\Upsilon = 3.5$) and
  the best-fit model without a black hole in NGC~3585.  The difference
  in $\chi^2$ is summed from the outermost region to the innermost for
  each axis.  Positive $\Delta\chi^2$ indicates that the model with a black
  hole is preferred.  The total difference in $\chi^2$ is different
  from the value listed in Table~\ref{t:masses} because this
  difference is between the two individual models rather than the
  marginalized results.  Most of the difference comes from the inner
  1\arcsec.}
\label{f:n3585cumdc}
\end{figure}

The $\chi^2$ contours in the $\mbh$--$\Upsilon$ plane are plotted in
Figure~\ref{f:n3585chi}.  The best-fitting model is able to reproduce
all of the major features in the velocity profile, and in order to
produce the increase in velocity dispersion seen toward the center, a
black hole is required.  The velocity profiles for the best-fit models
are shown in Figure~\ref{f:n3585spec}.  We marginalize over
mass-to-light ratio and take $\Delta\chi^2 = 1$ as our $1\sigma$
uncertainty to find a black hole mass of $\mbh = 3.4 {}_{-0.6}^{+1.5}
\times 10^{8}~\msun$.  For $\mbh = 0$, the marginalized $\chi^2$
increases from the minimum by 28.7, which rules out the absence of a
black hole at very high significance (better than $99.99\%$
confidence).  Marginalizing over black hole mass, we find a $V$-band
mass-to-light ratio $\Upsilon_V = 3.4 \pm 0.2$.  The best-fit model
with $\mbh = 3.4 \times 10^8~\msun$ and $\Upsilon = 3.5$ has $\chi^2 =
55.8$.  The difference between the best-fit model and the best-fit
model with $\mbh = 0$ is not obvious in Figure~\ref{f:n3585spec},
which shows the Gauss--Hermite moments of the LOSVDs.  For this reason,
we present a plot of the cumulative difference in $\chi^2$ between the
two models in Figure~\ref{f:n3585cumdc}.  The value for $\Delta\chi^2$
is different from the value shown in Table~\ref{t:masses}, which
reports the difference in \emph{marginalized} $\chi^2$ whereas
Figure~\ref{f:n3585cumdc} is the difference between two individual
models.  The best-fit model generally differs from the best-fit model
without a black hole by a larger amount.  The cumulative $\chi^2$ plot
shows that most of the difference comes from the central $\sim
1\arcsec$.

\subsection{NGC~3607}
\label{n3607}
\begin{figure}
\centering
%% \pdfbookmark[3]{Fig. 8: NGC 3607 chi-square contours}{pbf8}
\includegraphics[width=\figwidthtwo,angle=90]{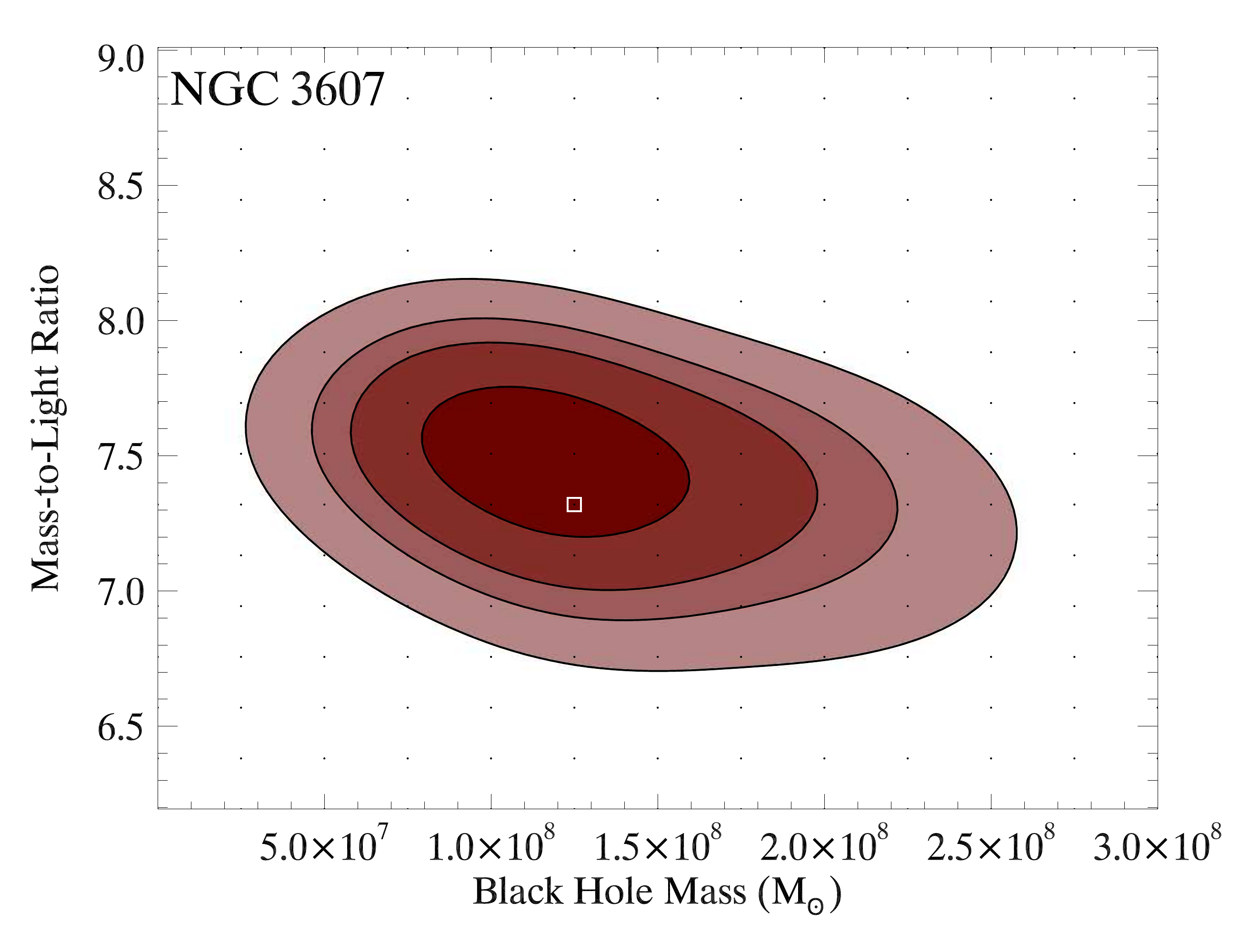}
\caption{Mass modeling $\chi^2$ contours for NGC~3607, assuming an
inclination of $51^\circ$.  Contours are the same as in
Figure~\ref{f:n3585chi}.  Contours have been smoothed for plotting.
The best-fit model has $M_{\mathrm{BH}} = 1.25\times10^8~\msun$ and
$\Upsilon = 7.3$.  Marginalizing over the other parameter, we find
$M_{\mathrm{BH}} = 1.2^{+0.4}_{-0.4}\times10^8~\msun$ and $\Upsilon =
7.5\pm 0.3$.}
\label{f:n3607chi}
\end{figure}

\begin{figure}
\centering
%% \pdfbookmark[3]{Fig. 9: NGC 3607 cumulative chi-square difference}{pbf9}
\includegraphics[width=\figwidth]{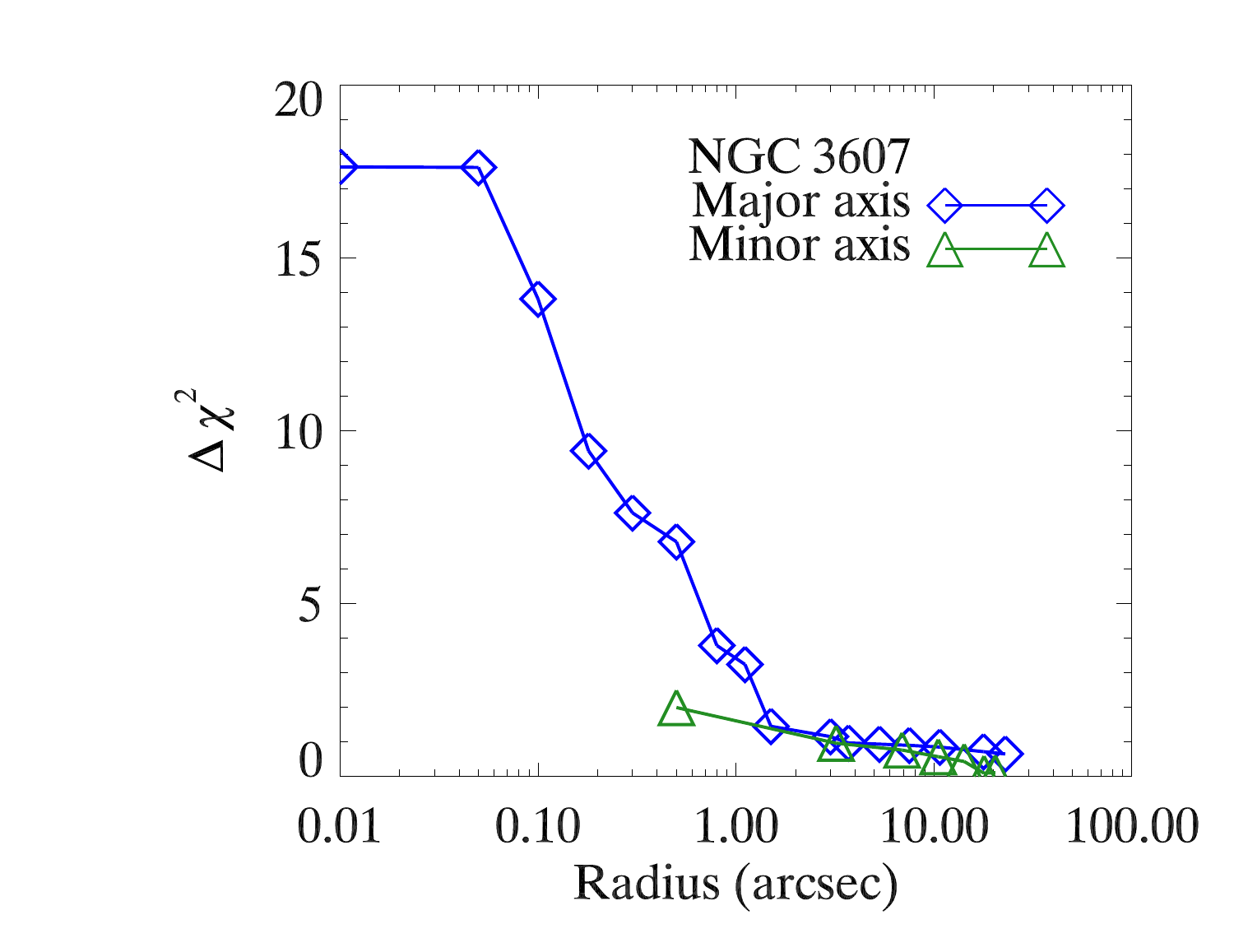}
\caption{Same as Figure~\ref{f:n3585cumdc} but for NGC~3607.  Positive
values of $\Delta\chi^2$ indicate preference for the best-fit model
($M_{\mathrm{BH}} = 1.25\times10^8~\msun$ and $\Upsilon = 7.3$).  Most
of the difference comes from the central 1\arcsec.}
\label{f:n3607cumdc}
\end{figure}

\object[NGC 3607]{NGC~3607} is an elliptical galaxy at a distance of
$19.9 \Mpc$ \citep{rc3,tonryetal01}.  The galaxy has a nearly opaque
dust disk toward the center \citep{laueretal05}, and the image
indicates a nearly face-on profile.  For this galaxy we assumed an
inclination angle of $51^\circ$, corresponding to a true axis ratio of
0.4. The dust obscures a large part of the bulge, but the nucleus is
still visible and is thus suitable for modeling.  The velocity
dispersion is roughly flat with radius, but there is an increased
rotation in the inner 0\farcs1, indicating a dark mass.  The effective
stellar velocity dispersion $\sigma_e = 229\kms$.  The $\chi^2$
contours in the $\mbh$-$\Upsilon$ plane are plotted in
Figure~\ref{f:n3607chi}.  The velocity profiles for the best-fit
models are shown in Figure~\ref{f:n3607spec}.  Marginalizing over
$\Upsilon$, we find a black hole mass of $\mbh = 1.2 {}_{-0.4}^{+0.4}
\times 10^{8}~\msun$.  For $\mbh = 0$, the marginalized $\chi^2$
increases from the minimum by 10.6, which rules out the absence of a
black hole at the $99.9\%$ confidence level.  Marginalizing over black
hole mass, we find a $V$-band mass-to-light ratio $\Upsilon_V = 7.5
\pm 0.3$.  The best-fit model with $\mbh = 1.25 \times 10^8~\msun$ and
$\Upsilon = 7.3$ has $\chi^2 = 69.12$.  Figure~\ref{f:n3607cumdc}
shows the cumulative $\chi^2$ as a function of radius, which indicates
that most of the difference comes from the central $\sim 1\arcsec$.

\subsection{NGC~3945}
\label{n3945}
At a distance of $19.9 \Mpc$, \object[NGC 3945]{NGC~3945} is an SB0 galaxy with a 
pseudobulge \citep{rc3,tonryetal01,laueretal07,kk04}.
This galaxy also contains both an inner bar system and an inner disk
\citep[e.g.,][]{2004A&A...415..941E,1999ApJ...521L..37E}.  The bars
may present problems four our axisymmetric code, which we discuss in
\S~\ref{concl}.
The velocity profile shows a rise in velocity dispersion toward the
center from $r=0\farcs2$, but this may also be interpreted as a slight
dip in velocity dispersion at $r\approx0\farcs2$--$0\farcs3$.  It has
an effective stellar velocity dispersion $\sigma_e = 192\kms$.  The
velocity profiles for the best-fit models are shown in
Figure~\ref{f:n3945spec}.

\begin{figure}
\centering
%% \pdfbookmark[3]{Fig. 10: NGC 3945 chi-square contours}{pbf10}
\includegraphics[width=\figwidthtwo,angle=90]{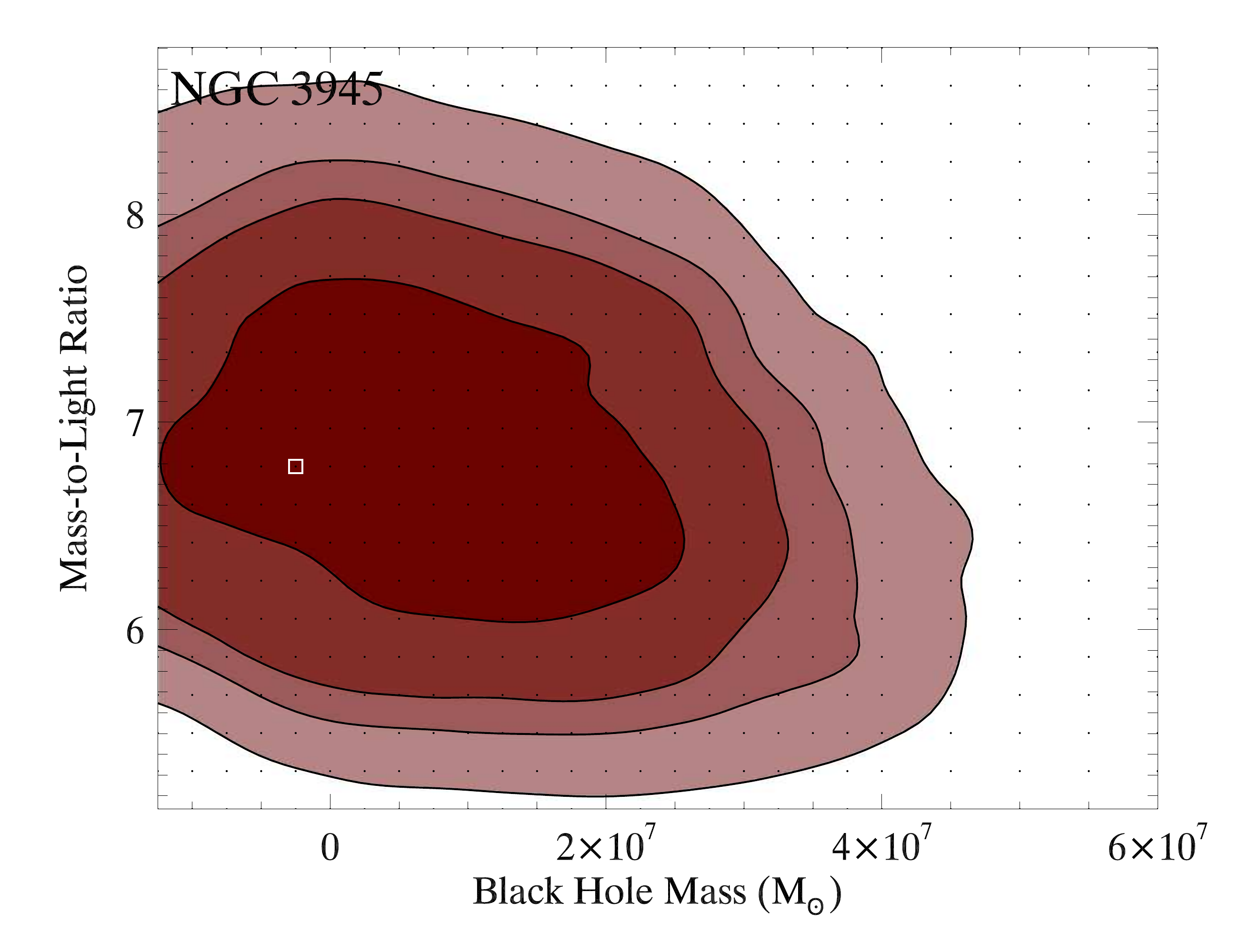}
\caption{Mass modeling $\chi^2$ contours for NGC~3945, assuming
edge-on inclination.  Contours are as in
Figure~\protect{\ref{f:n3585chi}}.  Contours have been smoothed for
plotting.  The best-fit model has $M_{\mathrm{BH}} =
-2.5 \times 10^6~\msun$ and $\Upsilon = 6.8$.  Marginalizing over the
other parameter we find $M_{\mathrm{BH}} = 9^{+17}_{-21}\times10^6~\msun$
and $\Upsilon = 6.6\pm 0.8$.}
\label{f:n3945chi}
\end{figure}

The $\chi^2$ contours in the $\mbh$-$\Upsilon$ plane are plotted in
Figure~\ref{f:n3945chi}.  The results of the kinematic modeling show
that this galaxy is consistent with no black hole at its center.
Marginalizing over $\Upsilon$, our estimates of the black hole mass
for any inclination do not exclude a black hole mass of zero at the
1$\sigma$ level: $\mbh = 9 {}_{-21}^{+17} \times 10^{6}~\msun$.  The
2$\sigma$ upper limit is $\mbh < 3.8 \times 10^7~\msun$, and the
3$\sigma$ upper limit is $\mbh < 5.1 \times 10^7~\msun$.  

Because $\mbh = 0$ was allowed for this galaxy, we included negative
black hole masses in our parameter space coverage.  This let us
consider the full extent of the 1$\sigma$ error distribution on the
low-mass side.  Our model allows a negative black hole mass as long as
the total mass inside the smallest pericenter of the orbit library is
positive.  In essence, this produces a delta function decrement to the
mass density.

\begin{figure}
\centering
%% \pdfbookmark[3]{Fig. 11: NGC 3945 cumulative chi-square difference}{pbf11}
\includegraphics[width=\figwidth]{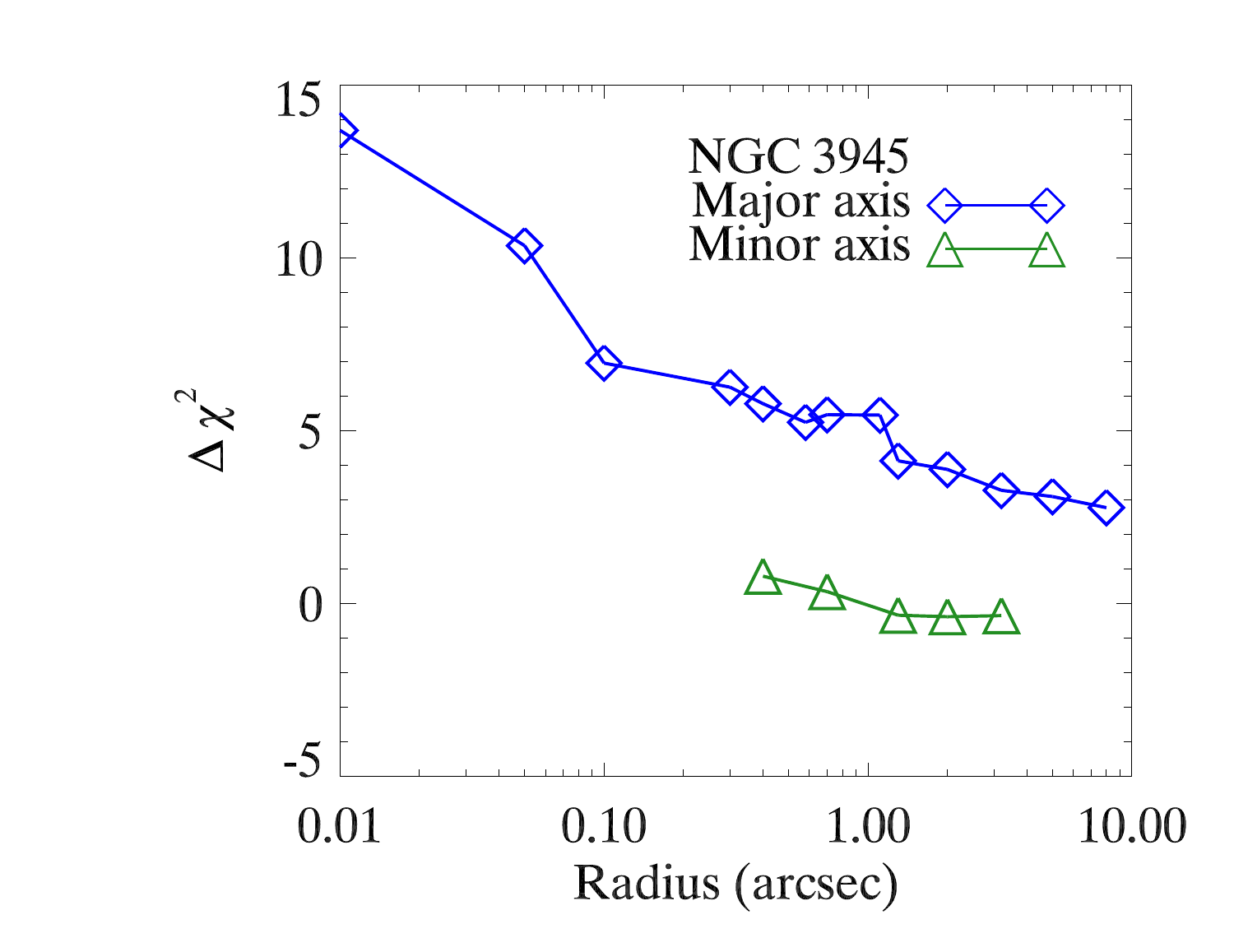}
\caption{Same as Figure~\ref{f:n3585cumdc} but for NGC~3945.  Positive
values of $\Delta\chi^2$ indicate preference for the best-fit model
($M_{\mathrm{BH}} = 0$ and $\Upsilon = 8.4$) compared to the best-fit
model with $\mbh = 1.0\times10^8~\msun$, the mass for a marginally
resolved sphere of influence.  Most of the difference comes from the
central 0\farcs1.}
\label{f:n3945cumdc}
\end{figure}

The sphere of influence of the black holes of the mass we find is
below the resolution limit of our data.  The black hole mass for
$\rinfres = 0.5$ is $\mbh = D\;\theta_{\mathrm{res}} {\sigma}^2 / G
= 9.7 \times 10^7~\msun$, where $D$ is the distance to the galaxy and
$\theta_{\mathrm{res}} = 0\farcs05$ is the spatial resolution
limit of the spectra.  Such a black hole mass, however, is ruled out
by our modeling under our assumptions of constant mass-to-light ratio
and axisymmetry, as inside $0\farcs1$ it would produce an excess
velocity dispersion above that observed, especially in the central
STIS pixel.  Marginalizing over black hole mass, we find $\Upsilon_V =
6.6 \pm 0.8$.  The model with $\mbh = -2.5 \times 10^6~\msun$ and
$\Upsilon = 6.8$ has $\chi^2 = 45.4$.  Figure~\ref{f:n3945cumdc}
shows the cumulative $\chi^2$ as a function of radius, which indicates
that most of the difference comes from the central $\sim 0\farcs1$.

\subsection{NGC~4026}
\label{n4026}
\object[NGC 4026]{NGC~4026} is an S0 galaxy at a distance of
$15.6 \Mpc$ \citep{rc3,tonryetal01}.  Images show a very flattened
profile, indicating that edge-on models are appropriate for this
galaxy.  There is also a weak, cold stellar disk in the center.  The
spectra show a flat velocity dispersion from $r \approx 3$\arcsec to
$r \approx 0\farcs3$ of $\sigma \approx 160\kms$ Inside 0\farcs3, the
velocity dispersion increases quickly to $\sigma = 258\kms$ at the
center, a strong indication of increased mass-to-light ratio.  The
effective stellar velocity dispersion $\sigma_e = 180\kms$.

\begin{figure}
\centering
%% \pdfbookmark[3]{Fig. 12: NGC 4026 chi-square contours}{pbf12}
\includegraphics[width=\figwidthtwo,angle=90]{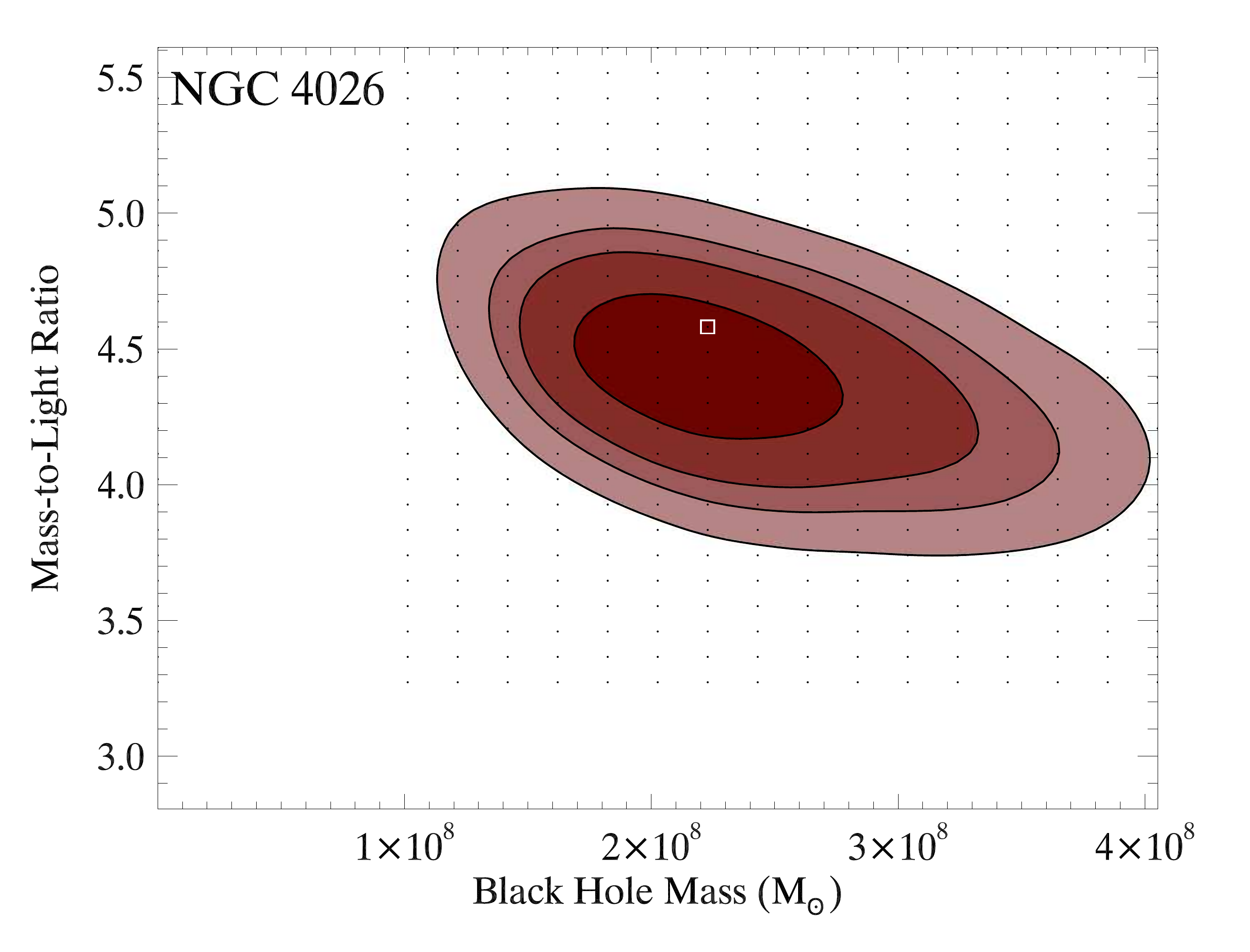}
\caption{Mass modeling $\chi^2$ contours for NGC~4026, assuming
edge-on inclination.  Contours are as in
Figure~\protect{\ref{f:n3585chi}}.  Contours have been smoothed for
plotting.  The best-fit model has $M_{\mathrm{BH}} =
2.2\times10^8~\msun$ and $\Upsilon = 4.6$.  Marginalizing over the
other parameter we find $M_{\mathrm{BH}} = 2.1^{+0.7}_{-0.4}\times10^8~\msun$
and $\Upsilon = 4.5\pm 0.3$.}
\label{f:n4026chi}
\end{figure}

\begin{figure}
\centering
%% \pdfbookmark[3]{Fig. 13: NGC 4026 cumulative chi-square difference}{pbf13}
\includegraphics[width=\figwidth]{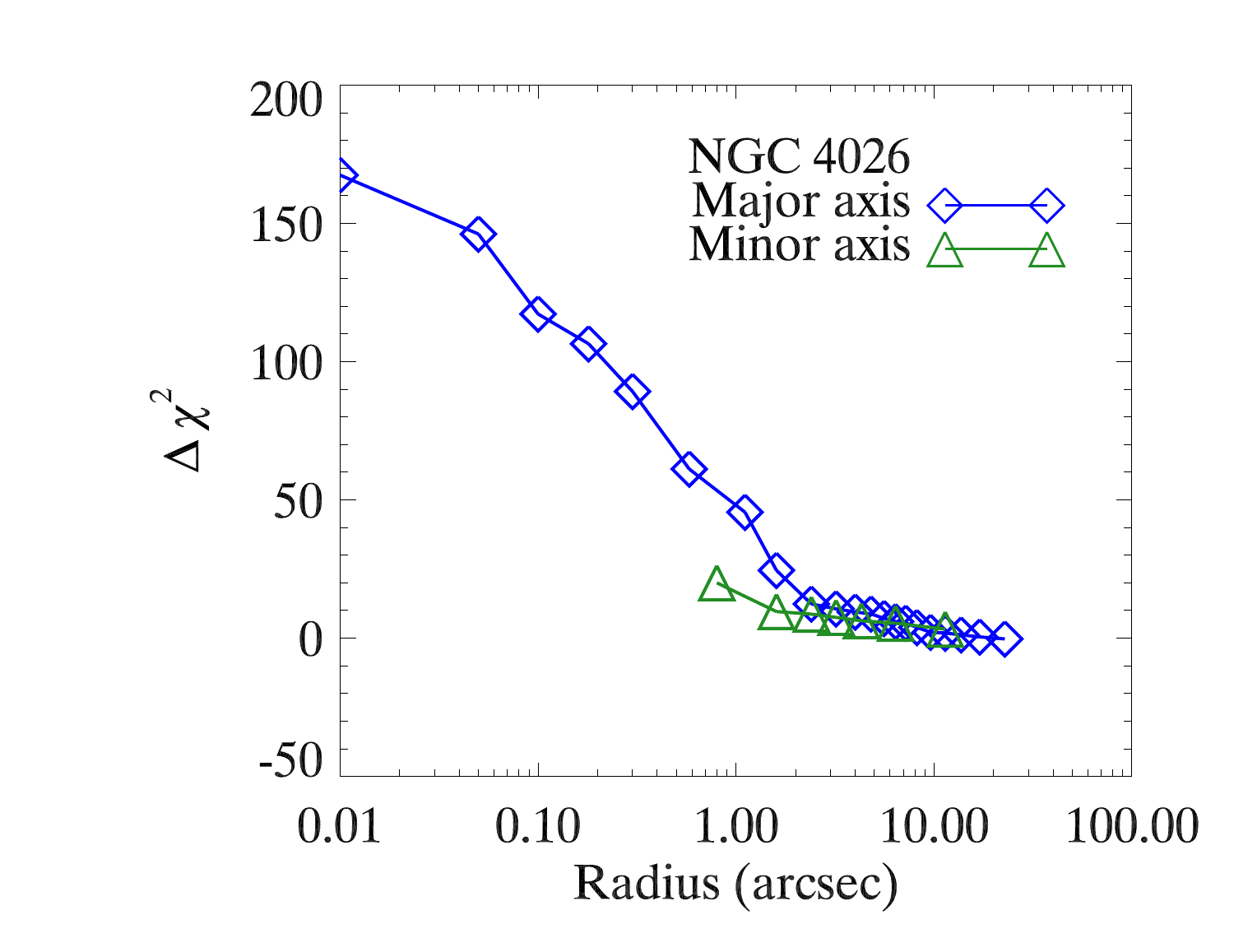}
\caption{Same as Figure~\ref{f:n3585cumdc} but for NGC~4026.  Positive
values of $\Delta\chi^2$ indicate preference for the best-fit model
($M_{\mathrm{BH}} = 2.2\times10^8~\msun$ and $\Upsilon = 4.6$).  Most
of the difference comes from the central 1\arcsec.}
\label{f:n4026cumdc}
\end{figure}

The $\chi^2$ contours in the $\mbh$-$\Upsilon$ plane are plotted in
Figure~\ref{f:n4026chi}.  The velocity profiles for the best-fit
models are shown in Figure~\ref{f:n4026spec}.  Marginalizing over
$\Upsilon$, we find a black hole mass of $\mbh = 2.1 {}_{-0.4}^{+0.7}
\times 10^{8}~\msun$.  For $\mbh = 0$, the marginalized $\chi^2$
increases from the minimum by 26.2.  Such an increase in
$\Delta\chi^2$ rules out the absence of a black hole at a confidence
level greater than $99.99\%$.  Marginalizing over black hole mass, we
find $\Upsilon_V = 4.5 \pm 0.3$.  The best-fit model with $\mbh = 2.2
\times 10^8~\msun$ and $\Upsilon = 4.6$ has $\chi^2 = 78.7$.
Figure~\ref{f:n4026cumdc} shows the cumulative $\chi^2$ as a function
of radius, which indicates that most of the difference comes from the
central $\sim 1\arcsec$.

\subsection{NGC~5576}
\label{n5576}
\begin{figure}
\centering
%% \pdfbookmark[3]{Fig. 14: NGC 5576 chi-square contours}{pbf14}
\includegraphics[width=\figwidthtwo,angle=90]{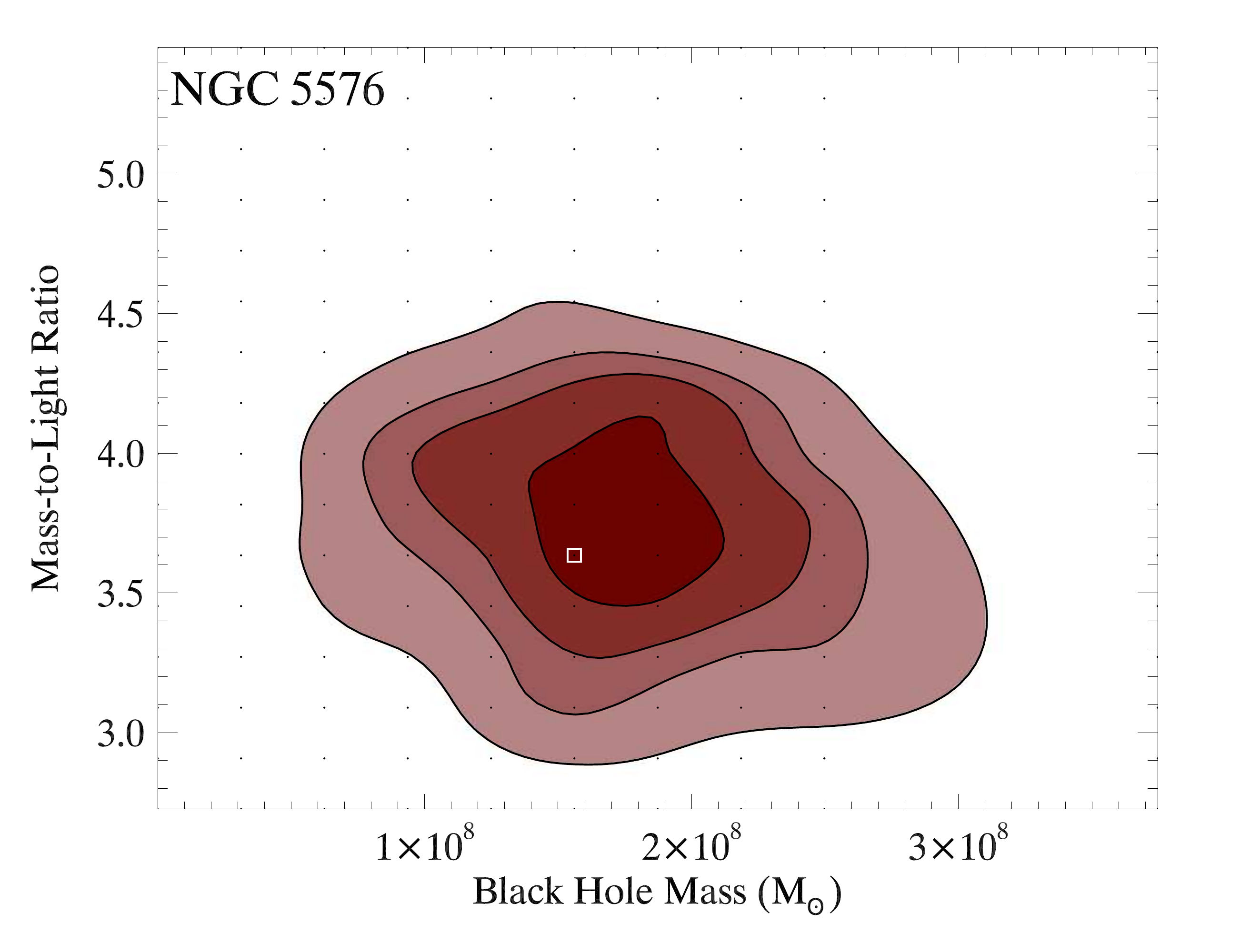}
\caption{Mass modeling $\chi^2$ contours for NGC~5576, assuming
edge-on inclination.  Contours are as in
Figure~\protect{\ref{f:n3585chi}}.  Contours have been smoothed for
plotting.  The best-fit model has $M_{\mathrm{BH}} =
1.6\times10^8~\msun$ and $\Upsilon = 3.6$.  Marginalizing over the
other parameter, we find $M_{\mathrm{BH}} = 1.8^{+0.3}_{-0.4}\times10^8~\msun$
and $\Upsilon = 3.7\pm 0.3$.}
\label{f:n5576chi}
\end{figure}

\begin{figure}
\centering
%% \pdfbookmark[3]{Fig. 15: NGC 5576 cumulative chi-square difference}{pbf15}
\includegraphics[width=\figwidth]{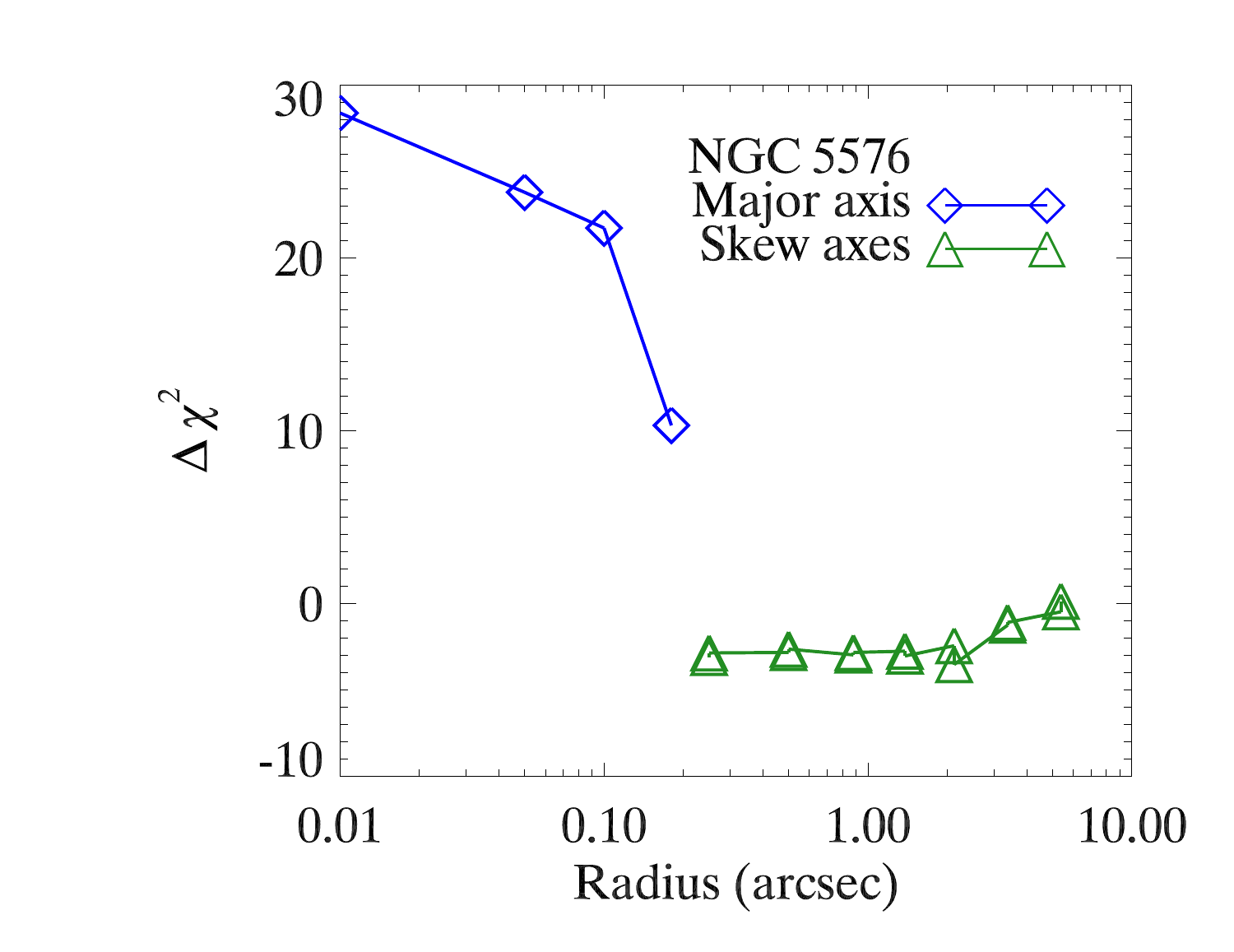}
\caption{Same as Figure~\ref{f:n3585cumdc} but for NGC~5576.  Positive
values of $\Delta\chi^2$ indicate preference for the best-fit model
($M_{\mathrm{BH}} = 1.6\times10^8~\msun$ and $\Upsilon = 3.6$).  Most
of the difference comes from the central 0\farcs2.}
\label{f:n5576cumdc}
\end{figure}

\object[NGC 5576]{NGC~5576} is an E3 radio galaxy at a distance of $27.1 \Mpc$
\citep{rc3,tonryetal01}.  The nucleus
is offset by $\sim0\farcs04$ from the center of the outer isophotes
\citep{laueretal05}.  Our ground-based spectroscopy qualitatively confirms 
this, showing that the central region is kinematically separate from the
outer regions.  The ground-based spectroscopy reveals an effective
stellar velocity dispersion of $\sigma_e = 183~\kms$.
Unlike the other modeling, for this galaxy we did not average both
sides of galaxy for the ground-based data. Instead, we included data
from both sides of the galaxy with the sign of the velocity
appropriately changed on one side.  Figure~\ref{f:n5576chi} shows the
$\chi^2$ contours from dynamical models.  The velocity profiles for
the best-fit models are shown in Figure~\ref{f:n5576spec}.
Marginalizing over $\Upsilon$, we find a black hole mass of $\mbh =
1.8 {}_{-0.4}^{+0.3} \times 10^{8}~\msun$.  At $\mbh = 0$, the
marginalized $\chi^2$ increases 15.5 above the minimum, indicating
that $\mbh = 0$ is ruled out at the $99.99\%$ confidence level.
Marginalizing over black hole mass, we nominally find $\Upsilon_V =
3.7 \pm 0.3$.  The best-fit model with $\mbh = 1.6 \times 10^8\msun$
and $\Upsilon = 3.6$ has $\chi^2 = 319.3$.  The value for $\chi^2$ is
much larger than it is for the other galaxies because (1) the
ground-based data come from 13 velocity bins instead of four
Gauss--Hermite moments, resulting in more constraints, and (2) there
are two sets of LOSVDs for each ground-based axis measurement: one
from each side of the galaxy.  Figure~\ref{f:n5576cumdc} shows the
cumulative $\chi^2$ as a function of radius, which indicates that most
of the difference comes from the central $\sim 0\farcs2$.


\section{Discussion and Summary}
\label{concl}
It is worth emphasizing that our modeling gives results for massive
dark objects at the centers of galaxies, which we call ``black
holes.''  Though the cumulative evidence in favor of these central
dark masses as black holes is strong, alternatives can only be ruled
out for the most highly resolved sources: the Galaxy, M31, and
NGC~4258 \citep[e.g.,][]{miller06}. Thus, while a large black hole is the most
astrophysically likely explanation of these central dark objects, they
do not strictly have to be black holes.

Our models assume a constant mass-to-light ratio for the stellar
component of the galaxy.  One possible source of systematic error may
come from the role that dark matter halos play.  Dark matter halos
likely increase the total mass-to-light ratio in the outer parts of
the galaxy but have less of an impact toward the center.  Thus, by
neglecting the dark matter halo, we may be overestimating the stellar
mass-to-light ratio at the center of the galaxy and, consequently,
underestimating the black hole mass.  Future models will incorporate
dark matter halos.

Another assumption of ours that may be violated is that of
axisymmetry.  Triaxiality has been addressed in other models
\citep{vandenboschetal08}.  In addition to triaxiality are bars, which
affect NGC~3945, a double-barred system.  While the primary bar is not
prominent inside of $15\arcsec$, where all of our kinematic data come
from, box orbits from bars can travel to the center.  Our modeling
code is axisymmetric and simply cannot model bars.  It is possible
that our modeling results, including the mass of the black hole, are
skewed by the bars.  If the bar is aligned mostly along the line of
sight, then the line-of-sight velocity would be higher than without
the bar.  The higher velocities, which would be observed, could be
misinterpreted as extra dark mass since the bar orbits would not be
accounted for in the model.  On the other hand, if the bar is aligned
mostly perpendicularly to the line of sight, the bar orbits would
contribute little to the line-of-sight velocity at the center, leading
to an underestimate in central dark mass.  One of the bars in NGC~3945
appears to lie in the plane of the sky, and one of the bars is at
least partially along the line of sight, assuming that they are
coplanar with the \emph{outer} disk.  Thus it is entirely possible
that the bars are leading to an incorrect inference of the black hole
mass, but it is not obvious whether it is skewed to a high or low
value.

\subsection{Anisotropy}
\label{anisotropy}
\begin{figure}
\centering
\includegraphics[width=\columnwidth]{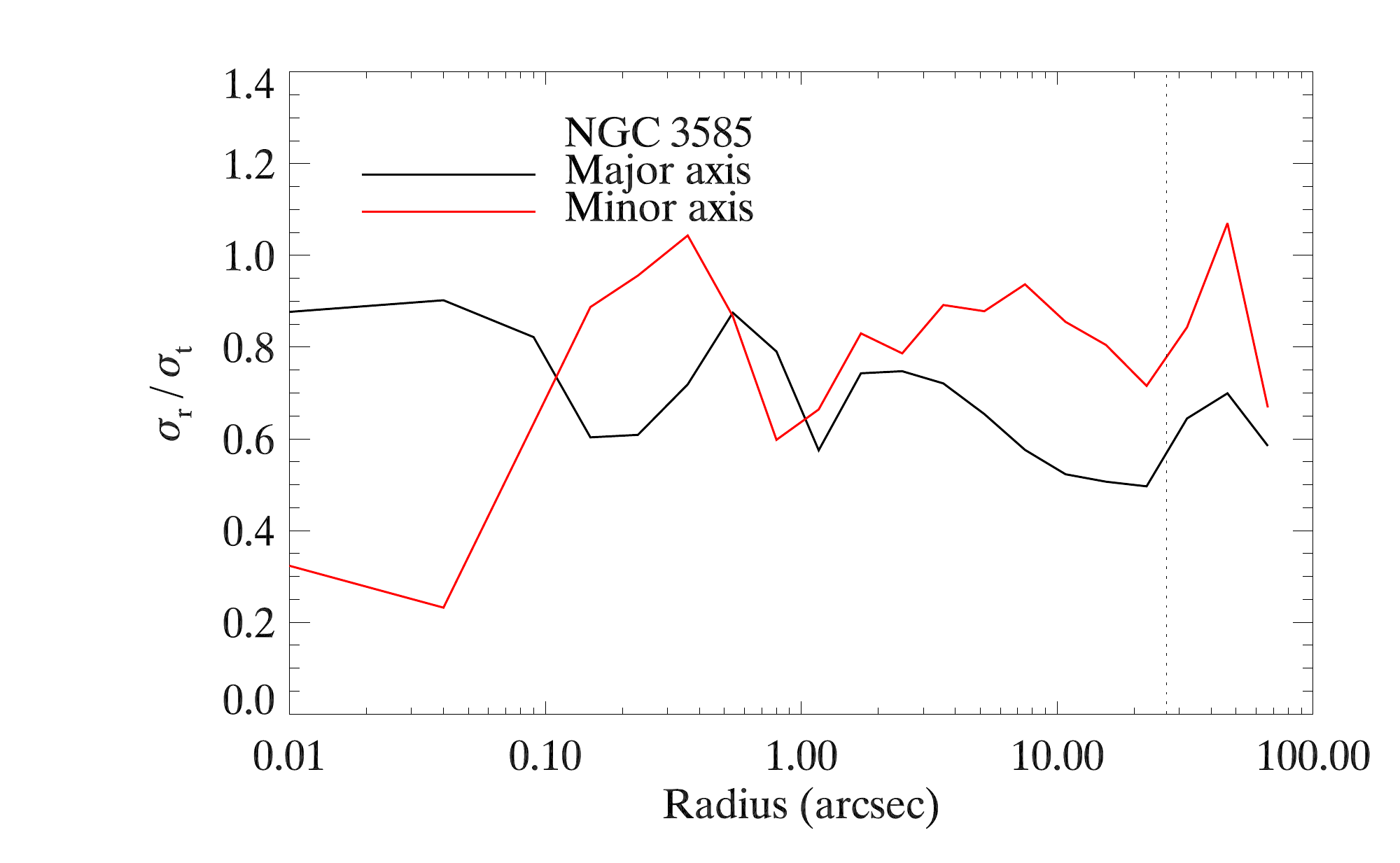}
\caption{Shape of the velocity dispersion tensor for NGC~3585 from the
best-fit model orbit solution.  The black line is along the major
axis, and the red line is along the minor axis.  The values for the
central part of the galaxy are plotted at a radius of 0\farcs01.  The
dotted line shows the radial extent of the ground-based spectroscopic
data.}
\label{f:n3585aniso}
\end{figure}

\begin{figure}
\centering
\includegraphics[width=\columnwidth]{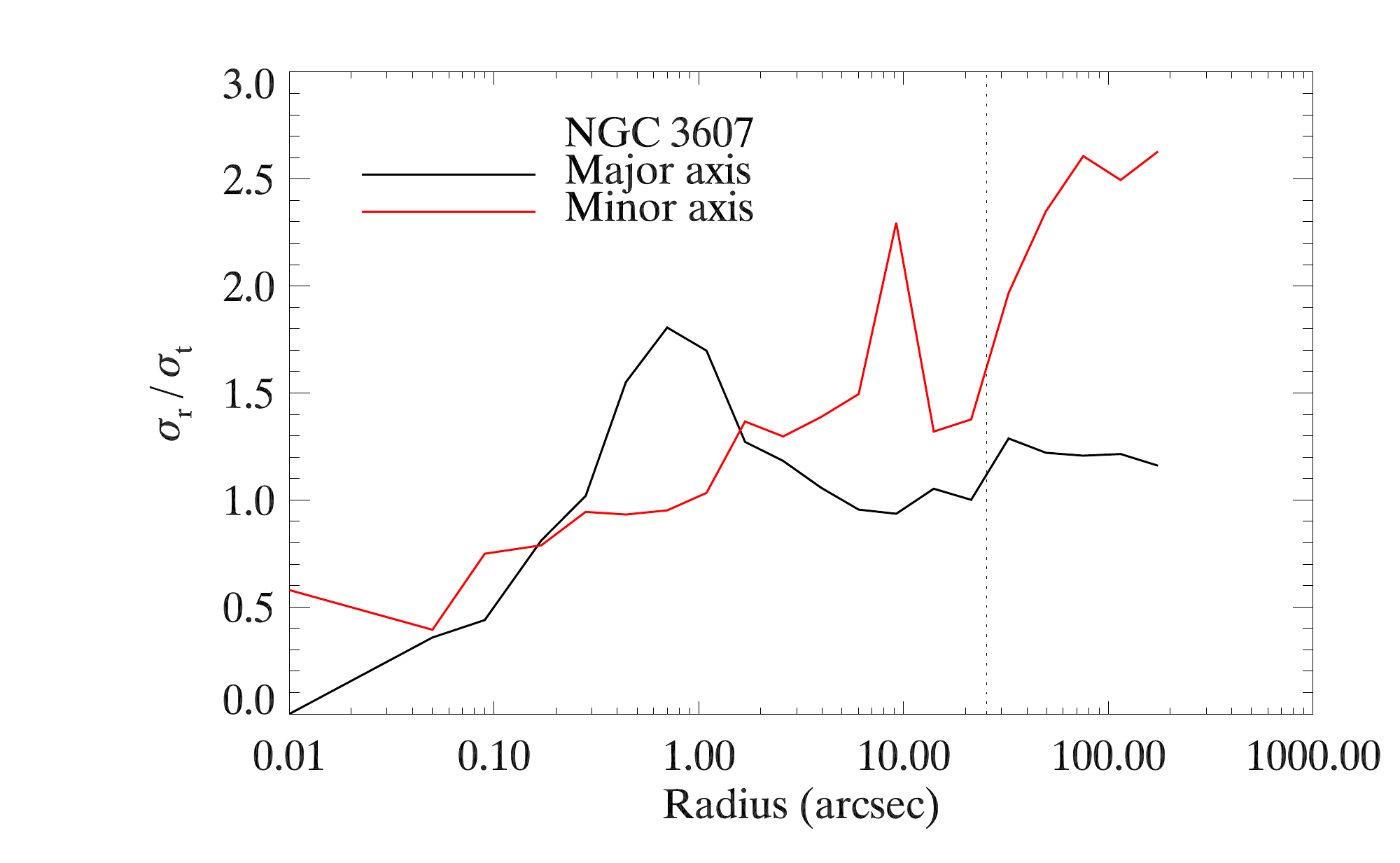}
\caption{Same as Figure~\ref{f:n3585aniso} but for NGC~3607.}
\label{f:n3607aniso}
\end{figure}

\begin{figure}
\centering
\includegraphics[width=\columnwidth]{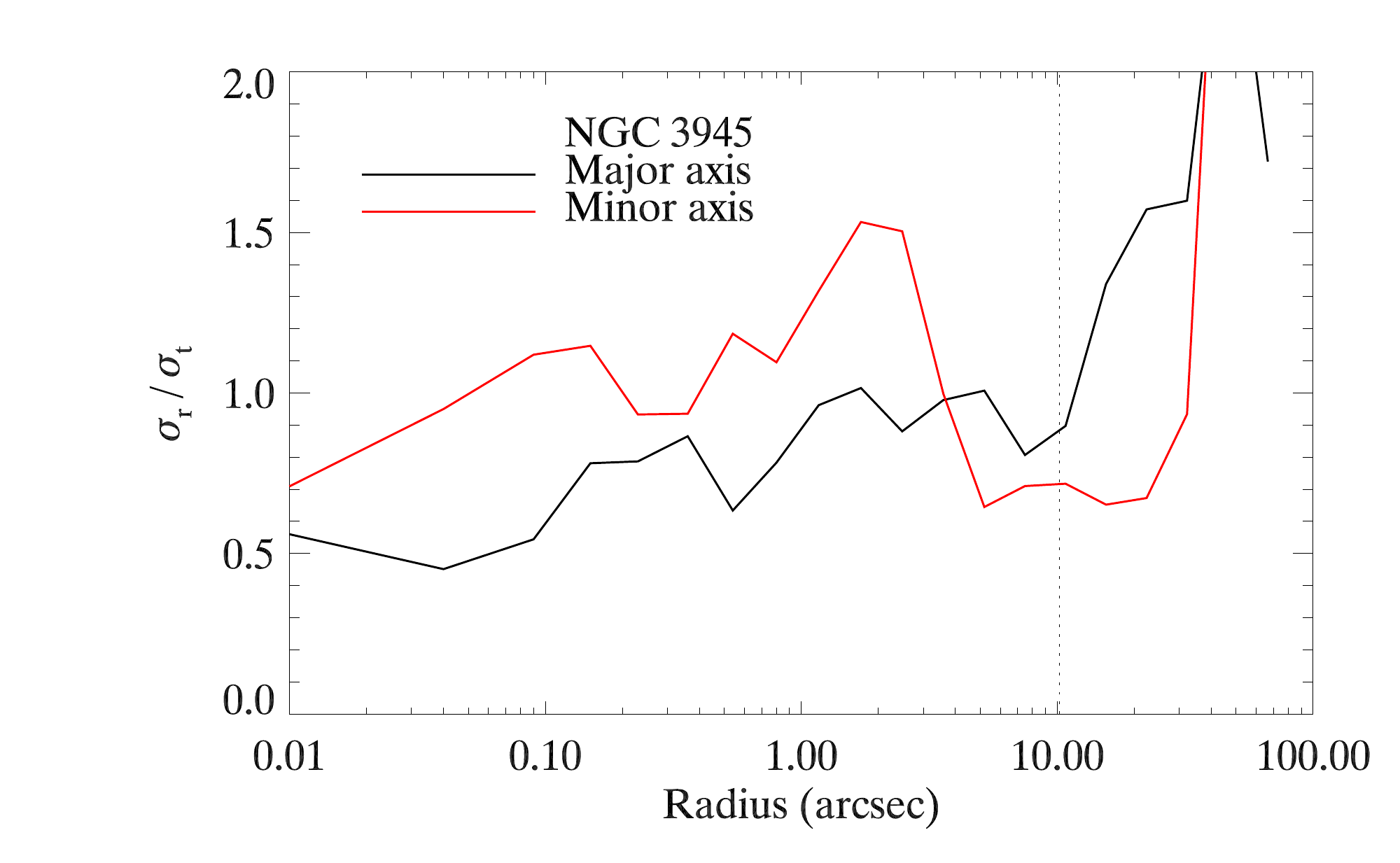}
\caption{Same as Figure~\ref{f:n3585aniso} but for NGC~3945.}
\label{f:n3945aniso}
\end{figure}

\begin{figure}
\centering
\includegraphics[width=\columnwidth]{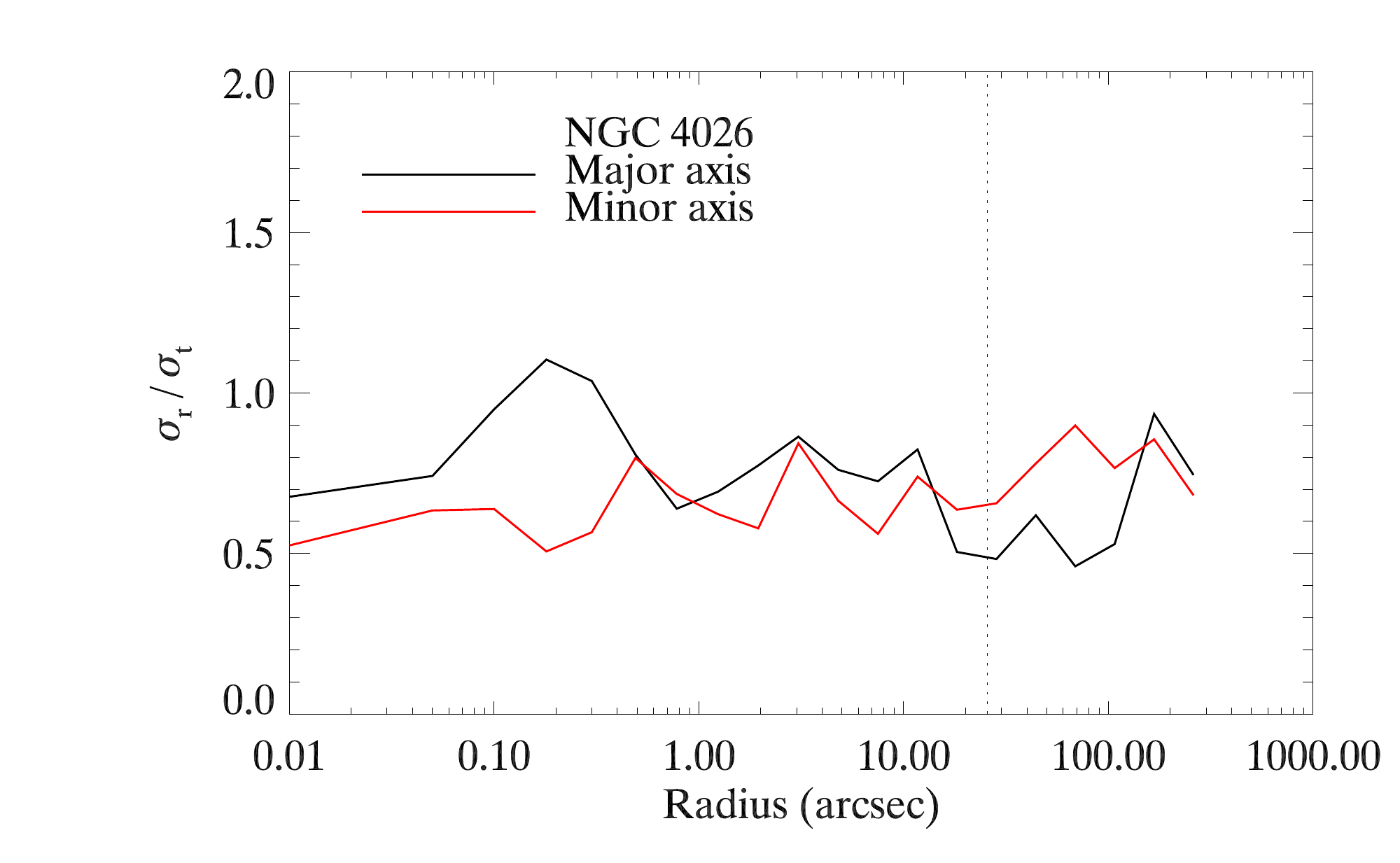}
\caption{Same as Figure~\ref{f:n3585aniso} but for NGC~4026.}
\label{f:n4026aniso}
\end{figure}

\begin{figure}
\centering
\includegraphics[width=\columnwidth]{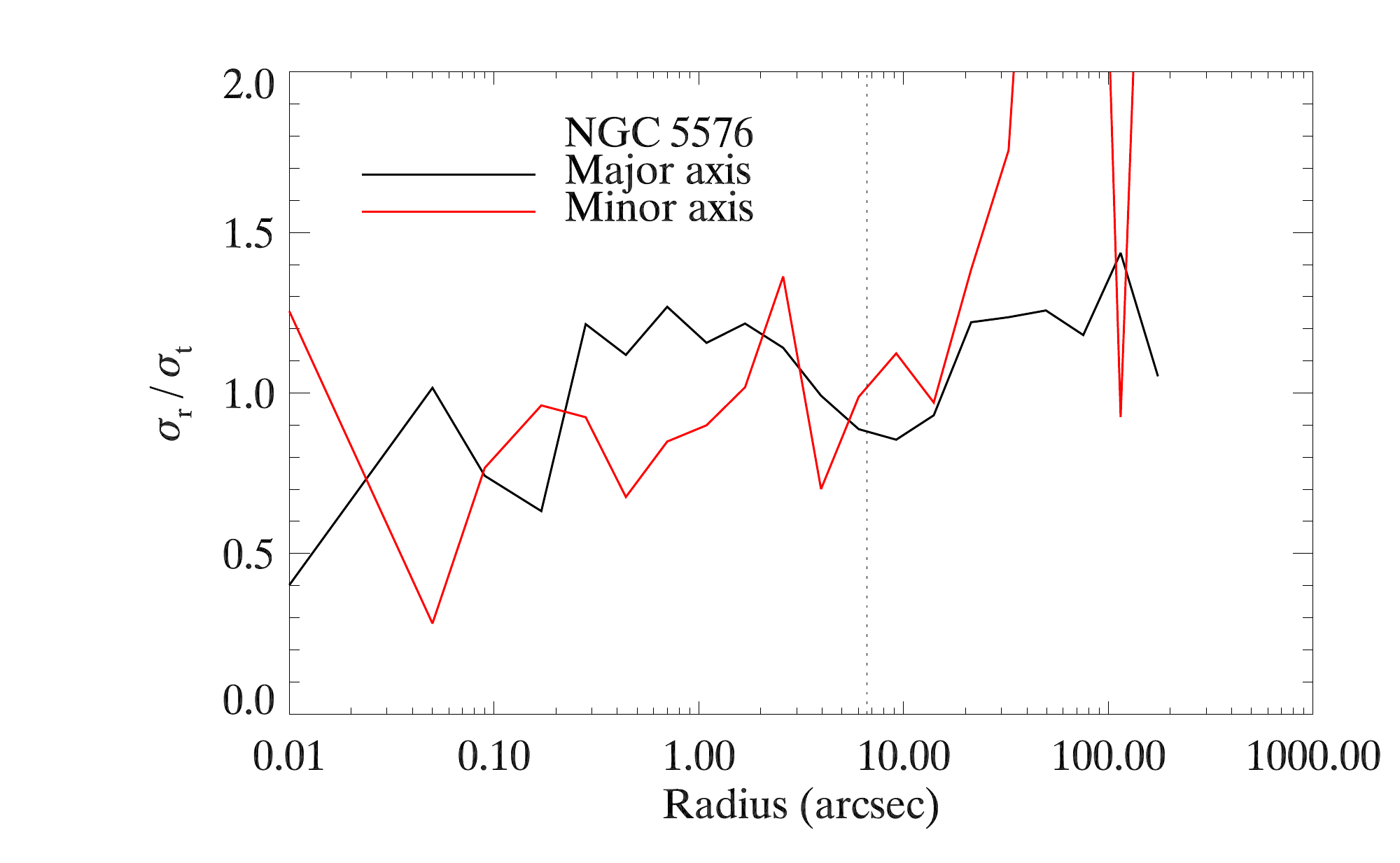}
\caption{Same as Figure~\ref{f:n3585aniso} but for NGC~5576.}
\label{f:n5576aniso}
\end{figure}

In Figures~\ref{f:n3585aniso}--\ref{f:n5576aniso}, we show the
velocity dispersion tensor for the best-fit model for each galaxy by
plotting the ratio of the radial velocity dispersion ($\sigma_r$) to
the tangential velocity dispersion, defined as $\sigma_t^2 \equiv 0.5
(\sigma_\theta^2 + \sigma_\phi^2)$ so that $\sigma_r / \sigma_t = 1$
for an isotropic distribution. Here, $\sigma_\phi$ is the second moment
of the azimuthal velocity relative to the systemic velocity rather
than relative to the mean rotational speed.  Uncertainties may be
estimated from the smoothness of the profiles \citep{gebhardtetal03}
to be $0.1$ to $0.3$.  All galaxies are dominated by tangential motion
at the center.

NGC~3585 (Figure~\ref{f:n3585aniso}) has an intermediate
surface-brightness profile and actually shows $\sigma_r / \sigma_t$
mildly increasing toward the center along the major axis, but with a
steep decrease inside of 0\farcs1 along the minor axis.  For almost
the entire range out to 23\arcsec, $\sigma_r / \sigma_t < 1$.  

The two core-profile galaxies, NGC~3607 (Figure~\ref{f:n3607aniso}) and
NGC~5576 (Figure~\ref{f:n5576aniso}), are both ellipticals and show
near-isotropic distributions in the outer regions, but both galaxies
are dominated by tangential motion at the center.  For NGC~3607, the
radial motion drops to almost zero.  NGC~3607 has a relatively strong
velocity gradient across the central 0\farcs3, and it has a strong
drop in the velocity dispersion in the center. The only way to
reproduce these observables is to have complete tangential anisotropy,
i.e., no radial orbits.  

The other two galaxies, NGC~3945 (Figure~\ref{f:n3945aniso}) and
NGC~4026 (Figure~\ref{f:n4026aniso}) are S0 galaxies with power-law
profiles.  For NGC~3945, outside of 10\arcsec, where the kinematic
data end, radial anisotropy dominates, but the uncertainties are large
here \citep{gebhardtetal03}.  Inside of 10\arcsec, the dispersion
along the major axis steadily decreases from a roughly isotropic value
to $\sigma_r / \sigma_t \approx 0.5$.  The dispersion along the minor
axis jumps from tangential to radial at $r \approx 3\arcsec$ and then
steadily decreases to $\sigma_r / \sigma_t \approx 0.7$.  NGC~4026
shows $\sigma_r / \sigma_t < 1$ almost everywhere.

\subsection{Demographics}
\label{demographics}
\begin{figure*}
\includegraphics[width=\figwidth]{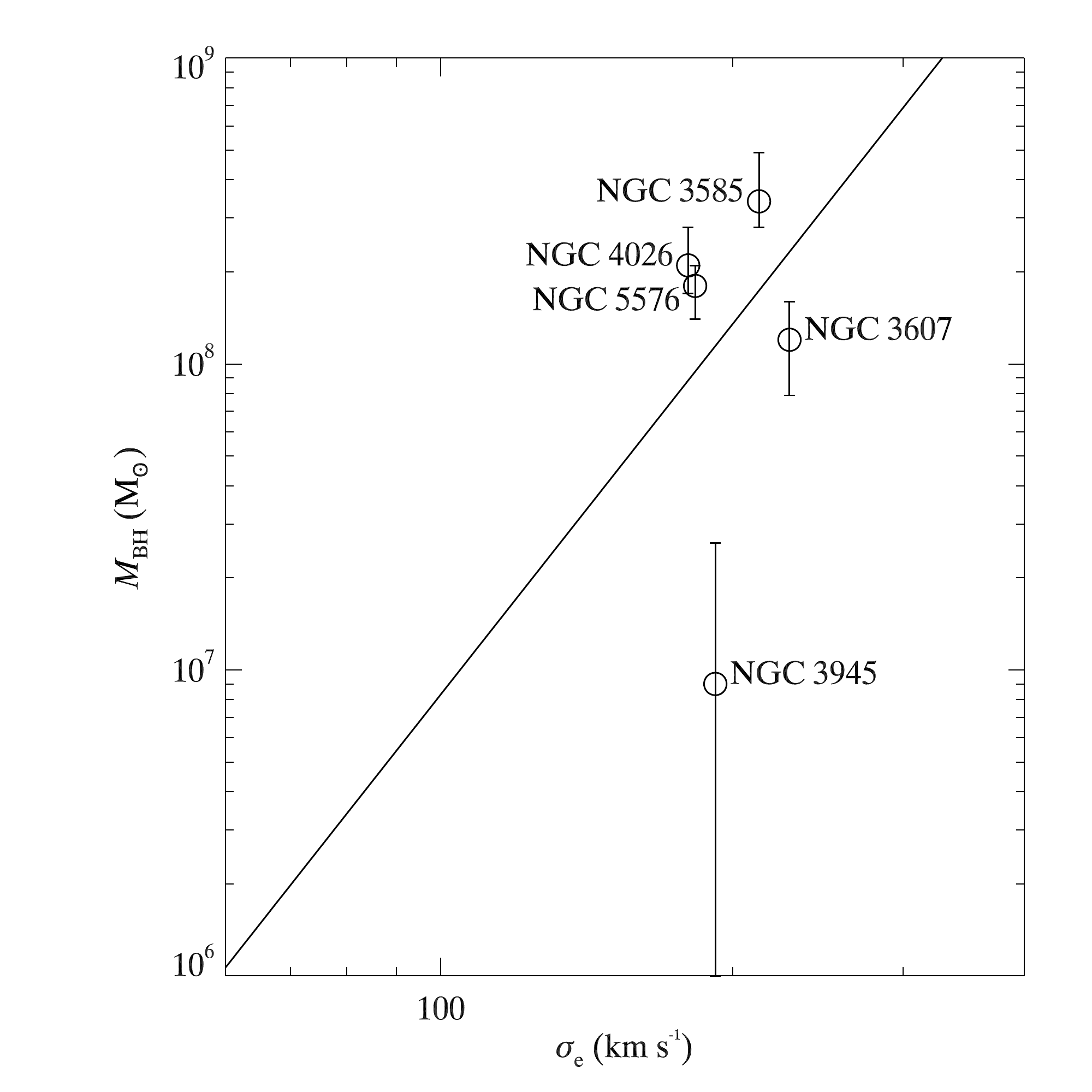}
\includegraphics[width=\figwidth]{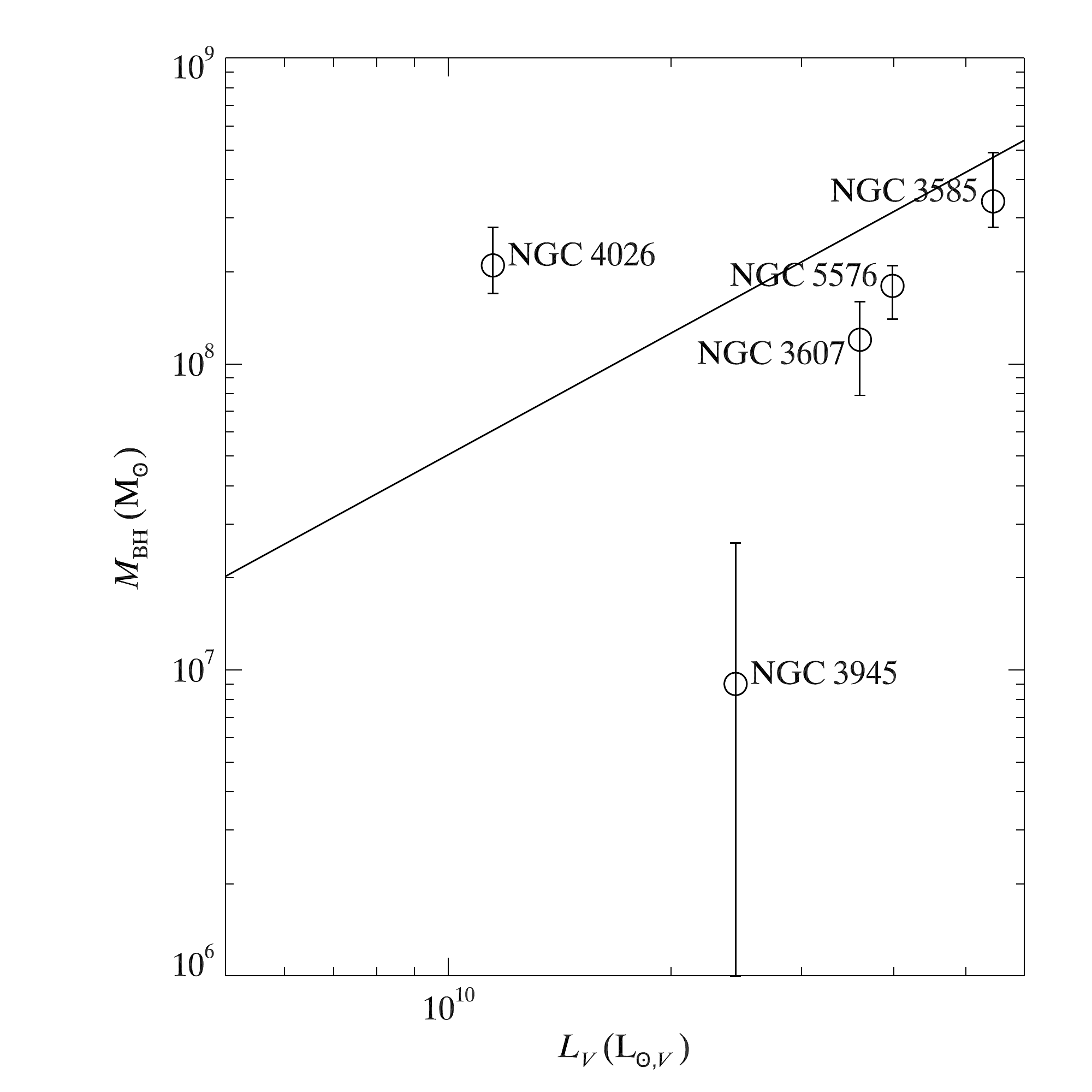}
\caption{Plots of black hole masses as functions of the host galaxies'
effective stellar velocity dispersion (\emph{left}) and luminosity
(\emph{right}).  The lines are the \msigma\ relation from
\protect{\citet{tremaineetal02}} (\emph{left}) and the \ml\ relation
from \protect{\citet{laueretal07}} (\emph{right}).  With the exception
of NGC~3945, all masses are found near the lines, but they are
\emph{all} inconsistent with the best-fit \msigma\ relation at the
1$\sigma$ level, and all but NGC~3585 are inconsistent with the \ml\
line.  Given the estimated scatter in the relations, most of the black
hole masses follow the relation.  NGC~3945, however, is significantly
below the relations.  Even the 3$\sigma$ upper limit to its mass
($M_{\mathrm{BH}} < 5.1 \times 10^7~\msun$) is more than a factor of 2
below the expected $1.1 \times 10^8~\msun$ from \msigma\ and $1.6
\times 10^8~\msun$ from \ml.}
\label{f:roughmrels}
\end{figure*}

We plot the masses found in \S~\ref{model} against $\sigma_e$ and
$L_V$ in Figure~\ref{f:roughmrels}, along with the \msigma\ and \ml\
relations from \citet{tremaineetal02} and \citet{laueretal07},
respectively.  With the exception of NGC~3585 in
Figure~\ref{f:roughmrels}(b), all of the black hole masses differ from
the values predicted by the scaling relations by at least 1$\sigma$.
\emph{This shows that our black hole mass measurements are precise
enough to probe the intrinsic scatter in these relations.}  The
measurement of the intrinsic scatter in these relations is addressed
by \citet{Gultekin_etal_2008b}.  If unaccounted systematic errors are
large, however, the residual scatter could be due to these.  Random
errors in distance are unlikely to be a large part of this as they are
typically 10\%, which is substantially smaller than the $\sim2$
deviation from the \msigma\ ridgeline.  For these particular
galaxies, inclination does not appear to make a significant difference
in black hole mass.  Systematic errors from triaxiality or bars,
however, are difficult to estimate and may contribute.

If one assumes an intrinsic scatter of 0.3~dex in the \msigma\
relation \citep[the maximum intrinsic scatter found
  by][]{tremaineetal02} and 0.5~dex in the \ml\ relation
\citep{laueretal07c}, all black hole masses are consistent with the
scaling relations except for NGC~3945, which is significantly below
both relations.  The black hole masses expected for NGC~3945 from the
\msigma\ and \ml\ relations are $1.1 \times 10^8~\msun$ and $1.6
\times 10^8~\msun$, respectively.  The 3$\sigma$ upper limit for
NGC~3945 is $\mbh < 5.1 \times 10^7~\msun$. Hence, while it not possible
to rule out the existence of a small black hole in NGC~3945, it does
not fall on the \msigma\ or \ml\ relations.

It is interesting to note that the single galaxy in our sample that is
consistent with having no black hole, NGC~3945, is a pseudobulge.
Comparing fits to the \msigma\ relation of pseudobulges with normal
bulges and ellipticals, \citet{hu08} concluded that pseudobulges have
systematically smaller black holes.  The small or absent black hole in
NGC~3945 is consistent with those findings.

\subsection{Summary}
\label{summary}
We conclude by summarizing the main results of this paper.  We
observed five early-type galaxies with high spatial resolution
kinematics from STIS, which we combined with WFPC2 photometry and
ground-based observations of photometry and kinematics.  We modeled
these data with three-integral, axisymmetric orbit models and found a
black hole mass consistent with zero and significantly below the
\msigma\ and \ml\ relations in one,
\begin{eqnarray*}
\mathrm{NGC~3945}&:&  ~M_{\mathrm{BH}} = 9^{+17}_{-21}\times10^{6}~\msun,\\
\end{eqnarray*}
though the presence of a double bar in this galaxy may present
problems for our axisymmetric code.  We find evidence for central
black holes in the remaining four:
\begin{eqnarray*}
\mathrm{NGC~3585}&:&  ~M_{\mathrm{BH}} = 3.4^{+1.5}_{-0.6}\times10^{8}~\msun, \\
\mathrm{NGC~3607}&:&  ~M_{\mathrm{BH}} = 1.2^{+0.4}_{-0.4}\times10^{8}~\msun, \\
\mathrm{NGC~4026}&:&  ~M_{\mathrm{BH}} = 2.1^{+0.7}_{-0.4}\times10^{8}~\msun,
\end{eqnarray*}
and
\begin{eqnarray*}
\mathrm{NGC~5576}&:& ~M_{\mathrm{BH}} = 1.8^{+0.3}_{-0.4}\times10^{8}~\msun. \\
\end{eqnarray*}
In all of these last four galaxies, the absence of a central dark 
object is ruled out to very high significance.

\acknowledgements We thank Louis Strolger for obtaining CCD images for
us during his run on the MDM 1.3 m as well as the anonymous referee.
K.G.\ thanks Marta Volonteri and Monica Valluri for helpful
discussions.  This work made use of the NASA's Astrophysics Data
System (ADS), and the NASA/IPAC Extragalactic Database (NED), which is
operated by the Jet Propulsion Laboratory, California Institute of
Technology, under contract with the National Aeronautics and Space
Administration. Financial support was provided by NASA/\emph{HST}
grants GO-5999, GO-6587, GO-6633, GO-7468, and GO-9107 from the Space
Telescope Science Institute, which is operated by AURA, Inc., under
NASA contract NAS 5-26555.

\bigskip
\bigskip
\bigskip
\bigskip
\bigskip
\bigskip
\bigskip
\bigskip
\bigskip

\eject
\bibliographystyle{apjads}
%%\bibliographystyle{apj}
%% \pdfbookmark[1]{References}{refbkmk}
\bibliography{gultekin}

\appendix 
\pdfbookmark[1]{Appendix}{appendixbkmk}
This Appendix gives tables of the original data used in this paper.
Details are available in \S~\ref{obs}.

\begin{deluxetable}{rr@{$\pm$}lr@{$\pm$}lr@{$\pm$}lr@{$\pm$}l}
  \footnotesize
  \tablecaption{Velocity Profile for NGC~3585}
  \tablehead{
     \colhead{Radius} &
     \multicolumn{2}{c}{$V$} &
     \multicolumn{2}{c}{$\sigma$} &
     \multicolumn{2}{c}{$h_3$} &
     \multicolumn{2}{c}{$h_4$} 
  }
  \startdata

0.00 & $-$8.1 & 31.7 & 281.2 & 29.6 & 0.030 & 0.069 & $-$0.034 & 0.061 \\
0.05 & 14.3 & 35.1 & 271.0 & 36.6 & $-$0.125 & 0.072 & $-$0.008 & 0.066 \\
0.10 & 73.2 & 29.2 & 260.9 & 33.5 & $-$0.156 & 0.067 & 0.044 & 0.080 \\
0.18 & 132.6 & 29.7 & 193.5 & 30.6 & $-$0.057 & 0.047 & $-$0.058 & 0.049 \\
0.30 & 147.7 & 29.5 & 211.2 & 24.6 & $-$0.181 & 0.066 & $-$0.010 & 0.054 \\
0.58 & 35.0 & 25.6 & 213.3 & 15.8 & $-$0.025 & 0.039 & $-$0.096 & 0.021 \\
1.12 & 150.7 & 25.7 & 185.8 & 22.5 & $-$0.109 & 0.060 & $-$0.015 & 0.064 \\
$-$0.05 & $-$42.1 & 26.9 & 244.4 & 35.2 & 0.106 & 0.077 & 0.094 & 0.070 \\
$-$0.10 & $-$112.4 & 29.6 & 185.2 & 71.8 & 0.016 & 0.115 & $-$0.020 & 0.151 \\
$-$0.18 & $-$88.8 & 28.9 & 197.2 & 21.1 & 0.024 & 0.035 & $-$0.060 & 0.036 \\
$-$0.30 & $-$13.3 & 46.5 & 248.6 & 34.4 & 0.185 & 0.071 & $-$0.045 & 0.063 \\
$-$0.58 & 12.3 & 37.9 & 198.9 & 22.2 & 0.147 & 0.054 & $-$0.038 & 0.037 \\
$-$1.12 & $-$104.4 & 32.0 & 239.6 & 27.5 & 0.320 & 0.091 & 0.218 & 0.107
\enddata
\label{t:n3585stisdata}
\tablecomments{Gauss--Hermite moments for velocity profiles derived 
from STIS data.  Radii are given in arcsec, first and second moments are 
given in units of $ \mathrm{km\;s^{-1}}$.}
\end{deluxetable}
\begin{deluxetable}{rr@{$\pm$}lr@{$\pm$}lr@{$\pm$}lr@{$\pm$}l}
  \footnotesize
  \tablecaption{Velocity Profile for NGC~3607}
  \tablehead{
     \colhead{Radius} &
     \multicolumn{2}{c}{$V$} &
     \multicolumn{2}{c}{$\sigma$} &
     \multicolumn{2}{c}{$h_3$} &
     \multicolumn{2}{c}{$h_4$} 
  }
  \startdata

0.00 & 13.2 & 22.8 & 180.5 & 22.1 & 0.077 & 0.068 & $-$0.007 & 0.063 \\
$-$0.05 & $-$36.1 & 29.0 & 227.3 & 35.1 & 0.119 & 0.086 & 0.069 & 0.111 \\
$-$0.10 & 8.9 & 26.3 & 151.4 & 27.0 & 0.000 & 0.034 & $-$0.065 & 0.066 \\
$-$0.18 & $-$59.9 & 49.4 & 216.9 & 39.8 & $-$0.167 & 0.096 & $-$0.021 & 0.102 \\
$-$0.30 & 39.1 & 33.9 & 206.9 & 29.5 & $-$0.071 & 0.091 & $-$0.040 & 0.098 \\
$-$0.58 & 22.8 & 31.4 & 197.2 & 32.3 & $-$0.003 & 0.079 & $-$0.050 & 0.072 \\
$-$1.12 & 49.2 & 27.5 & 160.2 & 25.7 & $-$0.011 & 0.076 & $-$0.031 & 0.084 \\
0.05 & $-$23.3 & 29.8 & 175.1 & 28.9 & 0.074 & 0.071 & $-$0.032 & 0.055 \\
0.10 & $-$67.4 & 34.2 & 197.6 & 44.4 & 0.117 & 0.105 & 0.031 & 0.120 \\
0.18 & $-$71.4 & 28.1 & 169.9 & 26.0 & 0.031 & 0.050 & $-$0.073 & 0.036 \\
0.30 & $-$21.7 & 87.7 & 251.7 & 41.6 & 0.222 & 0.178 & $-$0.002 & 0.397 \\
0.58 & $-$41.8 & 44.4 & 261.8 & 38.5 & $-$0.119 & 0.121 & $-$0.097 & 0.121 \\
1.12 & $-$27.7 & 38.5 & 197.3 & 36.3 & $-$0.005 & 0.083 & $-$0.073 & 0.073
\enddata
\label{t:n3607stisdata}
\tablecomments{Gauss--Hermite moments for velocity profiles derived 
from STIS data.  Radii are given in arcsec, first and second moments are 
given in units of $ \mathrm{km\;s^{-1}}$.}
\end{deluxetable}
\begin{deluxetable}{rr@{$\pm$}lr@{$\pm$}lr@{$\pm$}lr@{$\pm$}l}
  \footnotesize
  \tablecaption{Velocity Profile for NGC~3945}
  \tablehead{
     \colhead{Radius} &
     \multicolumn{2}{c}{$V$} &
     \multicolumn{2}{c}{$\sigma$} &
     \multicolumn{2}{c}{$h_3$} &
     \multicolumn{2}{c}{$h_4$} 
  }
  \startdata

0.00 & 16.9 & 10.1 & 170.7 & 8.1 & 0.032 & 0.036 & $-$0.073 & 0.030 \\
$-$0.05 & $-$14.0 & 11.4 & 173.3 & 9.8 & 0.058 & 0.035 & $-$0.100 & 0.029 \\
$-$0.10 & $-$36.0 & 16.9 & 164.4 & 16.5 & 0.146 & 0.044 & $-$0.002 & 0.045 \\
$-$0.18 & $-$46.6 & 16.2 & 148.0 & 18.9 & 0.111 & 0.059 & $-$0.015 & 0.057 \\
$-$0.30 & $-$24.0 & 18.5 & 125.4 & 22.5 & 0.093 & 0.058 & $-$0.036 & 0.062 \\
$-$0.58 & $-$100.1 & 19.0 & 138.9 & 24.2 & $-$0.038 & 0.073 & $-$0.037 & 0.079 \\
$-$1.12 & $-$29.4 & 20.5 & 138.5 & 20.5 & $-$0.035 & 0.057 & $-$0.063 & 0.075 \\
0.05 & 39.5 & 10.4 & 171.1 & 9.3 & 0.000 & 0.034 & $-$0.074 & 0.024 \\
0.10 & 12.8 & 11.8 & 158.6 & 9.9 & $-$0.021 & 0.039 & $-$0.075 & 0.036 \\
0.18 & 49.0 & 13.1 & 121.3 & 10.9 & $-$0.011 & 0.043 & $-$0.077 & 0.033 \\
0.30 & 61.3 & 18.0 & 140.5 & 18.0 & $-$0.053 & 0.069 & $-$0.025 & 0.048 \\
0.58 & 67.1 & 19.7 & 167.0 & 17.7 & $-$0.052 & 0.059 & $-$0.077 & 0.064 \\
1.12 & 101.0 & 16.1 & 153.6 & 16.1 & 0.062 & 0.064 & $-$0.009 & 0.055
\enddata
\label{t:n3945stisdata}
\tablecomments{Gauss--Hermite moments for velocity profiles derived 
from STIS data.  Radii are given in arcsec, first and second moments are 
given in units of $ \mathrm{km\;s^{-1}}$.}
\end{deluxetable}
\begin{deluxetable}{rr@{$\pm$}lr@{$\pm$}lr@{$\pm$}lr@{$\pm$}l}
  \footnotesize
  \tablecaption{Velocity Profile for NGC~4026}
  \tablehead{
     \colhead{Radius} &
     \multicolumn{2}{c}{$V$} &
     \multicolumn{2}{c}{$\sigma$} &
     \multicolumn{2}{c}{$h_3$} &
     \multicolumn{2}{c}{$h_4$} 
  }
  \startdata

0.00 & $-$52.0 & 23.1 & 258.3 & 32.0 & $-$0.019 & 0.066 & 0.139 & 0.078 \\
0.05 & $-$60.0 & 26.7 & 197.9 & 27.0 & $-$0.033 & 0.059 & $-$0.001 & 0.050 \\
0.10 & $-$75.6 & 27.3 & 197.1 & 24.2 & 0.098 & 0.050 & $-$0.004 & 0.050 \\
0.18 & $-$100.6 & 27.0 & 178.8 & 30.6 & 0.069 & 0.059 & 0.010 & 0.060 \\
0.30 & $-$79.7 & 27.5 & 163.3 & 22.0 & 0.048 & 0.049 & $-$0.057 & 0.053 \\
0.58 & $-$123.2 & 23.2 & 176.8 & 21.6 & 0.018 & 0.056 & $-$0.066 & 0.029 \\
1.12 & $-$105.5 & 29.8 & 187.5 & 31.6 & 0.000 & 0.061 & $-$0.074 & 0.060 \\
$-$0.05 & $-$31.5 & 36.5 & 360.5 & 42.1 & 0.158 & 0.102 & 0.099 & 0.104 \\
$-$0.10 & 34.5 & 27.1 & 237.7 & 28.4 & $-$0.004 & 0.066 & 0.010 & 0.063 \\
$-$0.18 & 148.2 & 24.1 & 173.7 & 26.3 & 0.002 & 0.059 & $-$0.049 & 0.051 \\
$-$0.30 & 189.8 & 20.7 & 155.6 & 16.4 & $-$0.006 & 0.034 & $-$0.052 & 0.023 \\
$-$0.58 & 170.9 & 18.5 & 143.9 & 16.7 & 0.007 & 0.049 & $-$0.044 & 0.020 \\
$-$1.12 & 186.4 & 15.3 & 132.1 & 14.4 & $-$0.041 & 0.037 & $-$0.040 & 0.034
\enddata
\label{t:n4026stisdata}
\tablecomments{Gauss--Hermite moments for velocity profiles derived 
from STIS data.  Radii are given in arcsec, first and second moments are 
given in units of $ \mathrm{km\;s^{-1}}$.}
\end{deluxetable}
\begin{deluxetable}{rr@{$\pm$}lr@{$\pm$}lr@{$\pm$}lr@{$\pm$}l}
  \footnotesize
  \tablecaption{Velocity Profile for NGC~5576}
  \tablehead{
     \colhead{Radius} &
     \multicolumn{2}{c}{$V$} &
     \multicolumn{2}{c}{$\sigma$} &
     \multicolumn{2}{c}{$h_3$} &
     \multicolumn{2}{c}{$h_4$} 
  }
  \startdata

0.00 & $-$1.2 & 17.3 & 226.4 & 18.3 & 0.088 & 0.054 & 0.059 & 0.054 \\
$-$0.05 & 17.7 & 15.7 & 217.7 & 22.6 & 0.057 & 0.051 & 0.039 & 0.048 \\
$-$0.10 & $-$66.7 & 25.4 & 183.6 & 85.9 & $-$0.016 & 0.147 & $-$0.035 & 0.173 \\
$-$0.18 & $-$38.0 & 18.5 & 202.1 & 22.6 & $-$0.117 & 0.049 & 0.041 & 0.034 \\
$-$0.30 & $-$17.6 & 20.0 & 218.8 & 15.1 & $-$0.156 & 0.048 & 0.019 & 0.042 \\
0.05 & $-$1.5 & 18.7 & 215.8 & 20.9 & 0.000 & 0.000 & 0.000 & 0.000
\enddata
\label{t:n5576stisdata}
\tablecomments{Gauss--Hermite moments for velocity profiles derived 
from STIS data.  Radii are given in arcsec, first and second moments are 
given in units of $ \mathrm{km\;s^{-1}}$.}
\end{deluxetable}

\begin{deluxetable}{lrr@{$\pm$}lr@{$\pm$}lr@{$\pm$}lr@{$\pm$}l}
  \footnotesize
  \tablecaption{Ground-Based Velocity Profile for NGC~3945}
  \tablehead{
     \colhead{Axis} &
     \colhead{Radius} &
     \multicolumn{2}{c}{$V$} &
     \multicolumn{2}{c}{$\sigma$} &
     \multicolumn{2}{c}{$h_3$} &
     \multicolumn{2}{c}{$h_4$} 
  }
  \startdata

Major & 0.00 & 0.0 & 0.0 & 199.9 & 2.7 & 0.016 & 0.003 & 0.147 & 0.147 \\
Major & 0.37 & 19.3 & 4.6 & 176.4 & 2.5 & 0.003 & 0.010 & 0.068 & 0.109 \\
Major & 0.74 & 42.5 & 11.0 & 180.9 & 3.5 & $-$0.051 & 0.010 & 0.065 & 0.096 \\
Major & 1.30 & 52.9 & 13.3 & 180.5 & 5.8 & $-$0.052 & 0.026 & 0.070 & 0.083 \\
Major & 2.04 & 75.8 & 12.5 & 201.2 & 6.9 & $-$0.054 & 0.049 & 0.093 & 0.115 \\
Major & 3.15 & 119.2 & 14.0 & 176.0 & 14.5 & $-$0.062 & 0.044 & 0.184 & 0.199 \\
Major & 5.01 & 154.1 & 9.2 & 175.9 & 7.9 & $-$0.121 & 0.014 & 0.215 & 0.223 \\
Major & 7.98 & 163.3 & 21.5 & 286.5 & 122.9 & $-$0.071 & 0.036 & 0.311 & 0.330 \\
Minor & 0.00 & 0.0 & 0.0 & 183.9 & 1.7 & $-$0.059 & 0.006 & 0.043 & 0.043 \\
Minor & 0.37 & 21.7 & 14.1 & 176.3 & 8.4 & $-$0.042 & 0.036 & 0.070 & 0.092 \\
Minor & 0.74 & $-$2.7 & 1.7 & 193.5 & 6.0 & $-$0.025 & 0.036 & 0.093 & 0.106 \\
Minor & 1.30 & 2.1 & 4.6 & 178.3 & 5.4 & $-$0.027 & 0.049 & 0.073 & 0.092 \\
Minor & 2.04 & $-$2.4 & 17.8 & 161.7 & 2.2 & 0.004 & 0.072 & 0.089 & 0.090 \\
Minor & 3.15 & $-$6.3 & 11.4 & 142.9 & 6.4 & $-$0.024 & 0.059 & 0.180 & 0.199
\enddata
\label{t:n3945groundspectra}
\tablecomments{Gauss--Hermite moments for velocity profiles derived 
from MDM data.  Radii are given in arcsec, first and second moments are 
given in units of $ \mathrm{km\;s^{-1}}$.}
\end{deluxetable}
\begin{deluxetable}{lrr@{$\pm$}lr@{$\pm$}lr@{$\pm$}lr@{$\pm$}l}
  \footnotesize
  \tablecaption{Ground-Based Velocity Profile for NGC~5576}
  \tablehead{
     \colhead{Axis} &
     \colhead{Radius} &
     \multicolumn{2}{c}{$V$} &
     \multicolumn{2}{c}{$\sigma$} &
     \multicolumn{2}{c}{$h_3$} &
     \multicolumn{2}{c}{$h_4$} 
  }
  \startdata

Major & 0.00 & 3.1 & 3.5 & 188.8 & 4.6 & 0.027 & 0.021 & $-$0.023 & 0.028 \\
Major & 0.25 & $-$1.0 & 3.2 & 192.9 & 4.9 & $-$0.018 & 0.020 & $-$0.015 & 0.020 \\
Major & 0.50 & 2.1 & 3.3 & 198.4 & 5.3 & $-$0.012 & 0.022 & $-$0.007 & 0.026 \\
Major & 0.88 & 4.3 & 3.7 & 191.3 & 5.8 & 0.013 & 0.020 & $-$0.004 & 0.023 \\
Major & 1.38 & 8.9 & 4.6 & 184.6 & 5.6 & $-$0.080 & 0.021 & $-$0.013 & 0.027 \\
Major & 2.12 & 10.0 & 4.5 & 166.8 & 7.9 & $-$0.045 & 0.023 & 0.118 & 0.022 \\
Major & 3.38 & 28.9 & 5.4 & 166.3 & 8.9 & $-$0.113 & 0.030 & 0.060 & 0.035 \\
Major & 5.38 & 16.7 & 7.8 & 162.6 & 12.3 & $-$0.041 & 0.053 & 0.006 & 0.052 \\
60$^\circ$ Skew & 0.00 & 2.0 & 3.6 & 191.6 & 5.6 & $-$0.029 & 0.021 & 0.011 & 0.025 \\
60$^\circ$ Skew & 0.25 & 4.9 & 3.4 & 188.9 & 5.5 & $-$0.017 & 0.023 & $-$0.009 & 0.023 \\
60$^\circ$ Skew & 0.50 & 3.3 & 3.8 & 188.2 & 5.1 & $-$0.012 & 0.019 & $-$0.014 & 0.026 \\
60$^\circ$ Skew & 0.88 & 2.6 & 3.7 & 191.2 & 5.9 & $-$0.016 & 0.023 & 0.010 & 0.027 \\
60$^\circ$ Skew & 1.38 & 3.3 & 4.5 & 192.0 & 6.7 & $-$0.012 & 0.023 & 0.018 & 0.025 \\
60$^\circ$ Skew & 2.12 & 9.6 & 4.1 & 190.1 & 4.5 & $-$0.008 & 0.026 & $-$0.062 & 0.024 \\
60$^\circ$ Skew & 3.38 & 11.0 & 5.0 & 171.7 & 5.6 & $-$0.102 & 0.027 & $-$0.010 & 0.035 \\
60$^\circ$ Skew & 5.38 & 23.1 & 5.3 & 146.7 & 7.0 & $-$0.110 & 0.028 & 0.021 & 0.032 \\
60$^\circ$ Skew & $-$0.25 & 3.1 & 3.8 & 195.1 & 6.3 & 0.026 & 0.020 & 0.025 & 0.026 \\
60$^\circ$ Skew & $-$0.50 & 1.0 & 4.2 & 196.0 & 8.4 & 0.036 & 0.021 & 0.050 & 0.027 \\
60$^\circ$ Skew & $-$0.88 & 1.4 & 4.6 & 183.8 & 6.7 & 0.051 & 0.022 & 0.027 & 0.025 \\
60$^\circ$ Skew & $-$1.38 & 15.7 & 4.0 & 168.1 & 6.2 & 0.032 & 0.023 & $-$0.018 & 0.023 \\
60$^\circ$ Skew & $-$2.12 & 7.4 & 4.4 & 159.6 & 6.6 & 0.025 & 0.028 & $-$0.026 & 0.030 \\
60$^\circ$ Skew & $-$3.38 & 10.5 & 5.0 & 158.2 & 7.9 & 0.048 & 0.028 & 0.013 & 0.028 \\
60$^\circ$ Skew & $-$5.38 & 13.8 & 6.2 & 157.5 & 7.7 & $-$0.028 & 0.034 & 0.018 & 0.029
\enddata
\label{t:n5576groundspectra}
\tablecomments{Gauss--Hermite moments for velocity profiles derived 
from Magellan data.  Radii are given in arcsec, first and second moments are 
given in units of $ \mathrm{km\;s^{-1}}$.}
\end{deluxetable}
\begin{deluxetable}{rrrr}
  \footnotesize
  \tablecaption{Ground-Based Surface Brightness Profile for NGC~3945}
  \tablehead{
     \colhead{Radius} &
     \colhead{Surface Brightness} &
     \colhead{Ellipticiy} &
     \colhead{P.A.} 
  }
  \startdata

10.28 & 18.909 & 0.352 & 156.1 \\
12.10 & 19.226 & 0.328 & 155.8 \\
14.23 & 19.664 & 0.236 & 154.8 \\
16.75 & 20.052 & 0.140 & 154.7 \\
19.70 & 20.416 & 0.038 &  94.4 \\
23.18 & 20.591 & 0.134 &  73.7 \\
27.27 & 20.678 & 0.262 &  70.6 \\
32.08 & 20.929 & 0.300 &  70.1 \\
37.74 & 21.517 & 0.175 &  74.6 \\
44.40 & 21.995 & 0.100 & 145.3 \\
52.23 & 22.415 & 0.169 & 150.0 \\
61.45 & 22.928 & 0.089 & 164.3
\enddata
\label{t:n3945groundphotometry}
\tablecomments{Radius is given in units of arcsec.  Surface brightness
is {$V$}-band in units of magnitudes per square arcsec.  The third
column gives ellipticity, and the fourth column gives position angle
in degrees east of north.}
\end{deluxetable}

\label{lastpage}
\end{document}